\documentclass[apj]{emulateapj}
\shorttitle{Nonrelativistic parallel shocks}
\shortauthors{Niemiec et al.}
\usepackage[]{hyperref}
\hypersetup{
   bookmarksnumbered=true,%
   bookmarksopen=true,%
   colorlinks=true,%
   urlcolor=blue,%
   pdftitle={Nonrelativistic parallel shocks in unmagnetized and weakly magnetized plasmas},%
   pdfauthor={Jacek Niemiec, Martin Pohl, Antoine Bret, Volkmar Wieland}}
\newcommand{\mfp}{$\lambda_{\rm mfp}$}

\begin{document}

\title{Nonrelativistic parallel shocks in unmagnetized and weakly magnetized plasmas}

\author{Jacek Niemiec\altaffilmark{1}, Martin Pohl\altaffilmark{2,3}, Antoine
Bret\altaffilmark{4,5,6}, Volkmar Wieland\altaffilmark{2,3}}
\altaffiltext{1}{Instytut Fizyki J\c{a}drowej PAN, ul. Radzikowskiego 152,
 31-342 Krak\'{o}w, Poland}
\altaffiltext{2}{DESY, 15738 Zeuthen, Germany}
\altaffiltext{3}{Institute of Physics and Astronomy, University of
Potsdam, 14476 Potsdam, Germany}
\altaffiltext{4}{Harvard-Smithsonian Center for Astrophysics, 60 Garden Street,
MS-51 Cambridge, MA 02138, USA}
\altaffiltext{5}{ETSI Industriales, Universidad de Castilla-La Mancha, 13071
Ciudad Real, Spain}
\altaffiltext{6}{Instituto de Investigaciones Energ´eticas y Aplicaciones
Industriales, Campus Universitario de Ciudad Real, 13071 Ciudad Real, Spain}

\email{jacek.niemiec@ifj.edu.pl}

\begin{abstract}
We present results of 2D3V particle-in-cell simulations of non-relativistic plasma collisions with absent or parallel large-scale magnetic field for parameters applicable to the conditions at young supernova remnants. We study 
the collision of plasma slabs of different density, leading to two different
shocks and a contact discontinuity. Electron dynamics play an important role in the development of the system. While non-relativistic shocks 
in both unmagnetized and magnetized plasmas can be mediated by Weibel-type
instabilities, the efficiency 
of shock-formation processes is higher when a large-scale magnetic field is present.
The electron distributions {downstream of the forward and reverse shocks} 
are generally isotropic, whereas that
is not always the case for the ions. We do not see any significant evidence of
pre-acceleration, neither in the electron population nor in the ion
distribution. 
\end{abstract}


\section{INTRODUCTION}
Collisionless shocks in space environments are generally considered to be sites
of efficient particle acceleration. One of the main unsolved problems is how 
efficiently individual particles can be fed into a Fermi-type
acceleration process. There, the particles have a mean free path, \mfp, large
enough 
that they see the shock as a sharp discontinuity and 
assume a power-law spectrum \citep{1987PhR...154....1B}. By definition,
collisionless
shocks have a thickness of a few \mfp\ of the thermal ions. In particular for 
electrons, pre-acceleration appears to be required for injection into Fermi-type
shock acceleration.

The shock transition region harbors small-scale electric and magnetic turbulence
that dissipate the flow energy and may provide pre-acceleration
\citep{2009ApJ...699..990B}.
Studies of nonrelativistic shocks indicate that the structure and
dissipation mechanisms
at these shocks vary considerably depending on the upstream plasma parameters
and the angle between the magnetic field and the
shock normal, $\theta_{\rm Bn}$. Quasi-perpendicular shocks, 
$\theta_{\rm Bn}\ge 45^\circ$, have been
studied for many years \citep[e.g.,][]{1988Ap&SS.144..535P,1988ApJ...329L..29C,
2004SSRv..110..161L}, and the basic structure of such shocks with Alfv\'en Mach
number
$M_{\rm A}\le 10$ is well understood \citep[e.g.,][]{2007PhPl...14a2108B}.
However, the physics of
quasi-perpendicular shocks in the limit of high Alfv\'en Mach number, 
as applicable to supernova remnants (SNRs),
is still under debate, in particular the electron injection
problem \citep{2009ApJ...695..574U,2009JGRA..11403217L,2010PhRvL.104r1102A,
2010ApJ...721..828K,2011ApJ...733...63R,2012arXiv1204.6312M}.

In contrast, quasi-parallel shocks, $\theta_{\rm Bn}\le 45^\circ$, have 
been harder to understand.
Observations of magnetic-field-aligned ion beams and
large-amplitude low-frequency
magnetic fluctuations upstream of space-plasma shocks have incited interest
in electromagnetic ion-beam instabilities \citep{1984JGR....89..179G} as sources
of
the necessary dissipation at the shock, particle injection, and
subsequent scattering {into} a Fermi-type acceleration mechanism
\citep[e.g.,][]{1988JGR....93.9649Q}. Studies of the
shock transition revealed that the shock ramp, where the principal density jump
occurs, is not steady, but instead cyclicly re-forms such that the shock
on average maintains its correct speed 
\citep{1990JGR....9518821W,1991JGR....9617715K}. These effects have 
often been studied with
one-dimensional hybrid simulations \citep{2011PhPl...18b2302S} that did not
resolve electron and short-scale ion spatio-temporal scales.
Short {particle-in-cell (PIC)} simulations of mildly relativistic quasiparallel
shocks were performed by
\citet{2010A&A...509A..89D} and \citet{mur10}. In quasiparallel conditions,
Whistler and
Alfv\'en waves are excited that in their nonlinear evolution grow
into Short Large-Amplitude Magnetic-field Structures (SLAMS), which are 
supposed to be efficient electron accelerators 
\citep{1997A&A...322..696C,2008ApJ...680L.153S,2009ApJ...693.1494M}.

\citet{kato08} recently reported the formation of nonrelativistic shocks
in unmagnetized
electron-ion plasmas via a Weibel-type short-wave instability \citep{weibel}
for 
relative {flow} speeds between $0.2c$ and $0.75c$ {($c$ is the speed of
light)}. 
{The relativistic two-stream Weibel (filamentation) instability \citep{medv99}
is known to mediate relativistic-shock formation in collisionless unmagnetized
or 
weakly-magnetized electron-positron 
\citep[e.g.,][]{1998ApJ...498L.183K,silva03,nishik03,chang08} and electron-ion
\citep[e.g.,][]{1998JPSJ...67.1079K,fred04,2008ApJ...673L..39S} plasmas. 
It leads to the development
of current filaments and accompanying toroidal magnetic fields. The results by
\citet{kato08} \citep[see also][]{stroman}
demonstrate} that a pre-existing magnetic
field is not necessary {for the existence of shock-mediating instabilities in
non-relativistic
flows. They also show} 
that electron dynamics {is} relevant for shock formation in unmagnetized plasma,
which 
calls for a fully kinetic (PIC) modeling in addition to hybrid simulations that
can cover a longer time period and have been successfully employed for oblique
nonrelativistic shocks \citep[e.g.,][]{2012ApJ...744...67G}.

We have
embarked on very-large-scale kinetic plasma simulations of the formation of
unmagnetized or {weakly-magnetized} strictly parallel 
{($\theta_{\rm Bn}=0^\circ$)} nonrelativistic shocks,
their long-term evolution, {and particle acceleration}. 
To address the question whether or not the shock structure and 
the particle spectra ever reach a steady state, we {follow} the evolution of
the system longer than {was done in \citet{kato08}}.
We are interested in asymmetric flows, i.e.,
the collision of plasma slabs of different density, leading to two different
shocks and a contact
discontinuity (CD). The Mach number is set high enough that electrostatic shocks
will not be formed. We want to see whether
a parallel magnetic field renders shock formation more efficient.
In the simulations with magnetized plasma, the magnetic-field strength is
assumed 
sufficiently {low} for the formation of the {forward and reverse} shocks with
large 
Alfv\'enic Mach number,
$M_{\rm A}\gg 1$. {This, together with the assumption of sub-relativistic plasma
collision speed,} renders our simulations applicable to the conditions
at young SNRs.

{In contrast to a conducting-wall reflection method to set up the shock, used in
\citet{kato08},
our asymmetric slab-collision model is physically more accurate as it avoids
assuming the existence of an infinitely sharp CD
{  (in fact a grid-cell wide)} and allows us to properly
investigate the dynamics of the 
forward and reverse shocks for different stream-counterstream parameters and
magnetic field
configurations. 
{  Beam-plasma systems are susceptible to various instabilities: some electrostatic 
(e.g., two-stream or Buneman modes), 
other quasi-electrostatic \citep[e.g.,][]{bret10}, and still others electromagnetic 
(e.g., filamentation).} 
The unstable spectrum is thus at least two-dimensional. 
Which of these modes would grow fastest highly depends on the system parameters 
\citep{2009ApJ...699..990B}. This clearly demonstrates a need for studies using
setup like ours, since the most unstable modes excited in various settings may
result in completely different shock-generating plasma waves.} 

{Besides {demonstrating} the efficiency of the shock-formation processes
in the presence of the mean magnetic field,} a secondary advantage 
of restricting to a parallel or absent large-scale magnetic field is that
simulations can be set up cleaner.
PIC simulations are sensitive to electromagnetic transients that arise from 
the initial setup and propagate through the simulation box. In shock simulations
they can induce scattering turbulence and electron heating in the upstream
region of
the shock. Likewise, a perpendicular component of the homogeneous magnetic
field 
implies that in the collision region (or reflecting plane, if
that is used to set up the shock) the curl of either the magnetic field 
or the motional electric field is large, effectively acting as an
antenna and emitting an electromagnetic pulse.

{The simulation model and setup are described in Section \ref{setup}. The
results of the simulations are
presented in Section \ref{results}. We conclude with a summary and discussion in
Section \ref{concl}.} 

\section{SIMULATION SETUP}\label{setup}
In our simulations, two electron-ion plasma streams of different densities
collide with each other at relative speed $v_{\rm rel}=0.38\,c$. The density 
ratio between the \emph{dense} and the \emph{tenuous} plasma slab is $10$. 
The simulation frame of reference is the center-of-momentum frame of the 
two plasmas, {in which} the dense (left)
plasma stream moves to the right in $+x$-direction with velocity 
${\bf{v}}_{\rm L}=0.0354\,c\,{\bf{\hat{x}}}$ and the tenuous (right) stream
moves
to the left in
$-x$-direction at ${\bf{v}}_{\rm R}=-0.354\,c\,{\bf{\hat{x}}}$. For simplicity, 
we will refer to the electrons and ions of the dense plasma stream as
{\emph{dense electrons}} and {\emph{dense ions}}, and {as \emph{tenuous
electrons} and \emph{tenuous ions} for the dilute plasma}. The plasmas are
initially very 
cold; in the respective rest frames of the streams, the electrons have a thermal
distribution with 
$v_{e,th}=\sqrt{kT/m_e}=0.002\,c$ and are in thermal equilibrium with the ions.
We consider 
both unmagnetized and magnetized plasma conditions. In the latter case, the
plasma flow is 
aligned with the homogeneous magnetic field, $B_{0,x}$, {and thus
we restrict our study to the formation of parallel shocks.}
The mean magnetic-field strength is given by the
ratio of the electron cyclotron frequency $\Omega_e=eB_{0,x}/m_e$ to the
electron plasma 
frequency of the dense-plasma component,
$\omega_{pe,L}=\sqrt{e^2N_{e,L}/\epsilon_0m_e}$ 
($\epsilon_0$ is the vacuum permittivity and $N_{e,L}$ is the density of dense
electrons), 
which is $\Omega_e/\omega_{pe,L}=0.04$. 
Such a flow-aligned magnetic-field allows filamentation-like instabilities. As shown in, e.g., \citet{bret06}, for the parameters of the initial plasma slab collision considered here and taking into account counterstreaming electron beams only, a plasma magnetization larger than
$\Omega_e/\omega_{pe,L}\simeq \beta_{\rm rel}\sqrt{0.1 \gamma_{\rm rel}}\approx 0.125$
would be required to suppress purely transverse filamentation modes, whereas a complete stabilization of oblique modes would never be efficient.
With the assumed plasma magnetization of one-third of the suppression field, the filamentation instability growth rate is  reduced by 10\% only.
The simulation setup is thus designed to
be applicable to young SNRs, in which dense supernova
ejecta propagating with nonrelativistic velocity collide with a dilute
weakly-magnetized interstellar medium. 

{On account of} the slower rate of shock-forming instabilities in
nonrelativistic flows, our 
large-scale simulations of nonrelativistic plasma collisions have been performed
in two 
spatial dimensions but keeping three components of the particle velocities (a
2D3V model). 
The code used in this study is a 2D3V modified version of the relativistic
electromagnetic 
particle code TRISTAN with MPI-based parallelization \citep{buneman93,niem08}. 
We have performed extensive test simulations to formulate a numerical model that
best conserves energy and minimizes numerical self-heating to provide a cold
plasma beam 
that is sufficiently stable against numerical 
grid-Cherenkov {and other instabilities}. 
The model takes advantage of second-order particle shapes and 
uses a second-order FDTD field-solver with a weak Friedman filter
\citep{green04} to suppress small-scale noise. 
In comparison with the numerical model used in previous work
\citep[e.g.,][]{niem08,stroman},
that employed first-order particle shapes and a computationally-expensive
iterative algorithm
for digital filtering of electric currents, the current model allows to maintain
a 
satisfactorily low noise level with fewer particles per cell, thus providing an
increased overall performance of the code \citep[compare][]{fonseca08}.  

\begin{deluxetable}{lcccc}
\tablecaption{Parameters of the Main Simulation Runs. \label{t1}}
\tablewidth{8.6cm}
\tablehead{
Run & $m_i/m_e$ & $\Omega_e/\omega_{pe,L}$ & Grid & $t^{max}$ \\
    &  & & ($\lambda_{si}^2$)  & ($\omega_{pi}^{-1}$)}
\startdata
U1 & 50 & 0 & 6122.5$\times$37.8    & 4111 \\
U2 & 20 & 0 & 5531.7$\times$39.8    & 3803 \\
M1 & 50 & 0.04 & 6122.5$\times$37.8   & 4111  \\
M2 & 20 & 0.04 & 5531.7$\times$39.8    & 3803 
\enddata
\tablecomments{Parameters of the main simulation runs described
in this paper. Listed are: the ion-electron mass ratio $m_i/m_e$, plasma
magnetization,
maximum computational box size in units of the ion skin depth of the dense
plasma 
($\lambda_{si}=57.2$ for runs U1 and M1 with $m_i/m_e=50$ and
$\lambda_{si}=36.2$ for runs 
U2 and M2 with $m_i/m_e=20$), and the run duration in units of
$\omega_{pi}^{-1}$.}
\end{deluxetable}

Table~\ref{t1} summarizes the parameters of the main simulation runs discussed
in this study.   
We use 5 particles per cell for each of the four plasma species in the active
grid and 
apply the splitting method for the tenuous plasma, that assigns statistical
weights to 
each particle to maintain the desired density ratio. Thus each ion particle can
be initialized 
at the same location as the corresponding electron, to ensure a vanishing
initial charge density. 
The electron skin depth of the dense plasma component
$\lambda_{se,L}=c/\omega_{pe,L}\simeq 8.1\Delta$, where $\Delta$ is the
grid-cell size. 
Because the characteristic length scales of the plasma ions should be well
contained in 
the simulation box, we can not use
a realistic proton-to-electron mass ratio. Here we base our discussion on the
results obtained 
for a reduced ion-electron mass ratio $m_i/m_e=50$ (runs U1 and M1 in Table 1). 
A slight dependence of the simulation results on the assumed mass ratio has been
observed in 
\citet{kato08}. To ascertain the effects of this parameter on the long-time 
evolution of the system we performed large-scale simulations {also}
for $m_i/m_e=20$ (runs U2 and M2);
the results of these simulations are summarized in Appendix A. 
For runs U1 and M1, referred here
as to the \emph{unmagnetized} and \emph{magnetized} simulation, respectively, 
the ion skin depth 
of the dense plasma {is}
$\lambda_{si}\equiv\lambda_{si,L}=c/\omega_{pi,L}\simeq 57.2\Delta$. 
Hereafter, $\lambda_{si}$ will be used as the unit of length and we take the
inverted 
ion plasma frequency of the dense component,
$\omega_{pi}^{-1}\equiv\omega_{pi,L}^{-1}$, as our unit of time.
The time step we use is $\delta t =
0.062\omega_{pe,L}^{-1}=0.0087\omega_{pi}^{-1}$. 

The simulations have been performed on a spatial grid in the $x-y$ plane with
radiating
boundary conditions in $x$ and periodic boundary conditions in $y$.
The transverse size of the computational grid used in runs U1 and M1 is
$L_y=37.8\,\lambda_{si}$ 
and the system evolution has been followed for $4111\,\omega_{pi}^{-1}$, 
i.e., $\sim2.9\times 10^4\,\omega_{pe,L}^{-1}$, which in simulation M1
corresponds
to $23.5\,\Omega_i^{-1}$, {where $\Omega_i$ is the ion cyclotron frequency.} 
The simulations thus require a very large grid size in plasma flow direction,
because the interpenetrating plasma streams occupy a continuously increasing
volume.
Our MPI-parallelized model partitions the simulation box into processor-domains
of equal size,
{and the number of 
processor-domains in the transverse direction is constant.}
To {limit}  the computational expense, in terms of both processor-usage 
and memory, an actual simulation is broken into several segments {
of different size in $x$, for which the number of 
processor-domains in flow direction is increased}. Thus, 
to complete an simulation with, e.g., $200,000$ time steps, succeeding runs 
with 1, 2, 4, and 6 216-processor domains each with $25,000 \times 10$ cells are
performed on
216, 432, 864, and 1296 cores, respectively. Particles are continuously added to
the system on 
both sides of the collision region with a so-called moving-injector. In this
method, the 
injection layer moves away from the interaction region at a distance that
preserves all the 
particles and fields generated in the collision region, but at the same time
does not allow 
a beam to travel a long distance without interaction with counterstreaming
particles, which 
further helps avoiding numerical grid-Cherenkov effects.  

Note that the parameters used in the main numerical experiments {were}
chosen following
an extended scan of the parameter space (e.g., number of particle per cell, the
electron skin 
depth, the magnetic field strength) performed with smaller-size simulations. The
results of the 
two selected production runs, U1 and M1, are thus representative of the physical
effects 
observed in the unmagnetized and weakly-magnetized system. The validity of the
two-dimensional 
simulations in capturing the essential physics of parallel shock formation
through the electromagnetic 
instabilities has been demonstrated in numerous {studies of}
relativistic and nonrelativistic shocks
\citep[e.g.,][]{2008ApJ...673L..39S,niem08,kato08}.     

The collision of the two plasma streams, as studied in this work, should in the
nonlinear stage 
result in the formation of a system of forward and reverse shocks separated by a
CD. The shock structures observed in our simulations can thus be
compared with 
the predictions of a hydrodynamic model. In analogy to the analysis presented 
in \citet{2009ApJ...698L..10N}, we can derive the CD speed, which in the
simulation frame is 
${\bf{\beta}}_{\rm CD}=-0.06\,{\bf{\hat{x}}}$. The CD moves because the system
is initially under momentum balance but not ram-pressure balance. 
The velocities of the forward and reverse shock depend on the adiabatic index
of the shocked plasma,
which is different for unmagnetized and magnetized plasmas in the simulations.
In the unmagnetized 
case, particle motion is confined to the simulation plane and the
nonrelativistic adiabatic 
index is $\Gamma_{\rm U}=2$. In the magnetized case, the magnetic field bends
the particle
trajectories out of the simulation plane. In effect, particles have three
degrees of freedom and 
the nonrelativistic adiabatic index is $\Gamma_{\rm M}=5/3$. 
{In the following we use subscripts ${\rm "U"}$ and ${\rm "M"}$ to denote quantities
measured in the unmagnetized and magnetized run, respectively. Subscripts ${\rm "R"}$ and ${\rm "L"}$
refer to the rest frame of the tenuous and dense plasma, respectively. 
These subscripts are dropped for quantities measured in the simulation frame.}

For an unmagnetized plasma collision, the steady-state speed of the forward and
the reverse 
shock should approach $\beta_{\rm fR,U}=0.43$ and $\beta_{\rm rL,U}=-0.14$, in
the rest frame of the tenuous and dense beam, respectively. In the simulation
frame, the shock velocities are  
$\beta_{\rm f,U}=0.09$ and $\beta_{\rm r,U}=-0.11$. The compression ratios at
each shock are predicted to be $N_{\rm fU,d}/N_{\rm fU,u}=2.9$ and $N_{\rm
rU,d}/N_{\rm rU,u}=3.01$. In the magnetized case, the simulation-frame
velocities of the forward and the reverse shocks
should be $\beta_{\rm f,M}=0.04$ and $\beta_{\rm r,M}=-0.09$, and the shock
compression
ratios $N_{\rm fM,d}/N_{\rm fM,u}=3.86$ and $N_{\rm rM,d}/N_{\rm rM,u}=4.02$,
respectively. 
The Alfv\'{e}n velocity{in the dense plasma is}
$v_{\rm A,L}=[B_{0,x}^2/\mu_0 (N_{e,L}m_e+N_{i,L}m_i)]^{1/2}=5.67\cdot
10^{-3}c$. 
In the rest frame of the tenuous plasma, the Alfv\'{e}n speed $v_{\rm
A,R}=1.85\cdot 10^{-2}c$. 
In the upstream rest frames of the tenuous and dense plasma the velocities of
the forward and
reverse shocks read $\beta_{\rm fR,M}=0.39$ and $\beta_{\rm rL,M}=-0.13$,
respectively. Thus, the
Alfv\'{e}nic Mach number of the forward shock $M_{\rm Af}=c\beta_{\rm
fR,M}/v_{\rm A,R}\simeq 21.1$ and that of the reverse shock $M_{\rm
Ar}=c\beta_{\rm rL,M}/v_{\rm A,L}\simeq 22.5$.
\footnote{A customary stated, the upstream flow Mach numbers read $M_{\rm
Af}=v_{\rm R}/v_{\rm A,R}\simeq 19.1$ and $M_{\rm Ar}=v_{\rm L}/v_{\rm
A,L}\simeq 6.3$.}

\section{SIMULATION RESULTS}\label{results}

In what follows, we present the results of our two main numerical experiments
for unmagnetized plasma conditions (run U1; Sec. \ref{sumag}) and for magnetized
plasma 
(run M1; Sec. \ref{smag}), {both of which use an} 
ion-electron mass ratio $m_i/m_e=50$. The results of 
two other large-scale runs, U2 
and M2, that assume a lower mass ratio $m_i/m_e=20$, are described in Appendix
A. 
The analysis of simulation data is also supported with additional test runs. 

To understand the nature of {various} unstable wave modes that appear in
the 
systems, we also perform a linear analysis.  
Given the complexity of the {situation} and the need to investigate the
{full spectrum
of unstable waves}, our linear calculations are restricted to the \emph{cold}
plasma regime. 
We thus write the cold fluid equations for each of the plasma components 
and derive the dielectric tensor $\mathbf{T}(\mathbf{k},\omega)$ 
adapting the Mathematica Notebook described in \cite{Notebook}.
For both unmagnetized plasma {and} a flow-aligned magnetic field,
we can {without loss of generality restrict the wavevector to the $x-y$
plane},
$\mathbf{k}=(k_x,k_y)$. 
The dispersion relation then reads $\det \mathbf{T}(\mathbf{k},\omega) =0$,
{from which we calculate the growth rates of unstable wave modes.}

\subsection{Unmagnetized Plasma}\label{sumag}
\subsubsection{Early-Stage Evolution}
In the first stage of the plasma collision, the interpenetration of dense and
tenuous streams 
quickly leads to current filamentation and {the} generation of transverse
(out-of-plane in our 
two-dimensional set-up) magnetic field in the collision
region. 
{The observed turbulent modes result from Weibel-like instabilities, and their
spectrum 
is dominated by modes with wave vectors oriented obliquely to the plasma flow
\citep{bret05}. This
is in agreement with our linear analysis in \cite{stroman}, that assumed similar
stream-counterstream 
relative velocity and density ratio.}   
The amplified magnetic 
field influences the motion of particles. The ion beams slightly decelerate in
bulk and develop 
a population of slow particles. At the same time the dilute ion beam, in which
filamentation is 
strongest, is heated. The bulk kinetic energy released by the decelerating ions
is converted to electron heating
in the electrostatic fields that accompany the ion filaments 
\citep[e.g.,][]{hed04,medv06,nishik06,2008ApJ...673L..39S}.
A fraction of the electrons of both streams is reflected away
from the collision 
region. The initially reflected electrons of the tenuous plasma constitute a
distinct hot beam 
that counterstreams against the
incoming particles of the tenuous stream. 

\begin{figure}[htb]
\epsscale{1.178}
\plotone{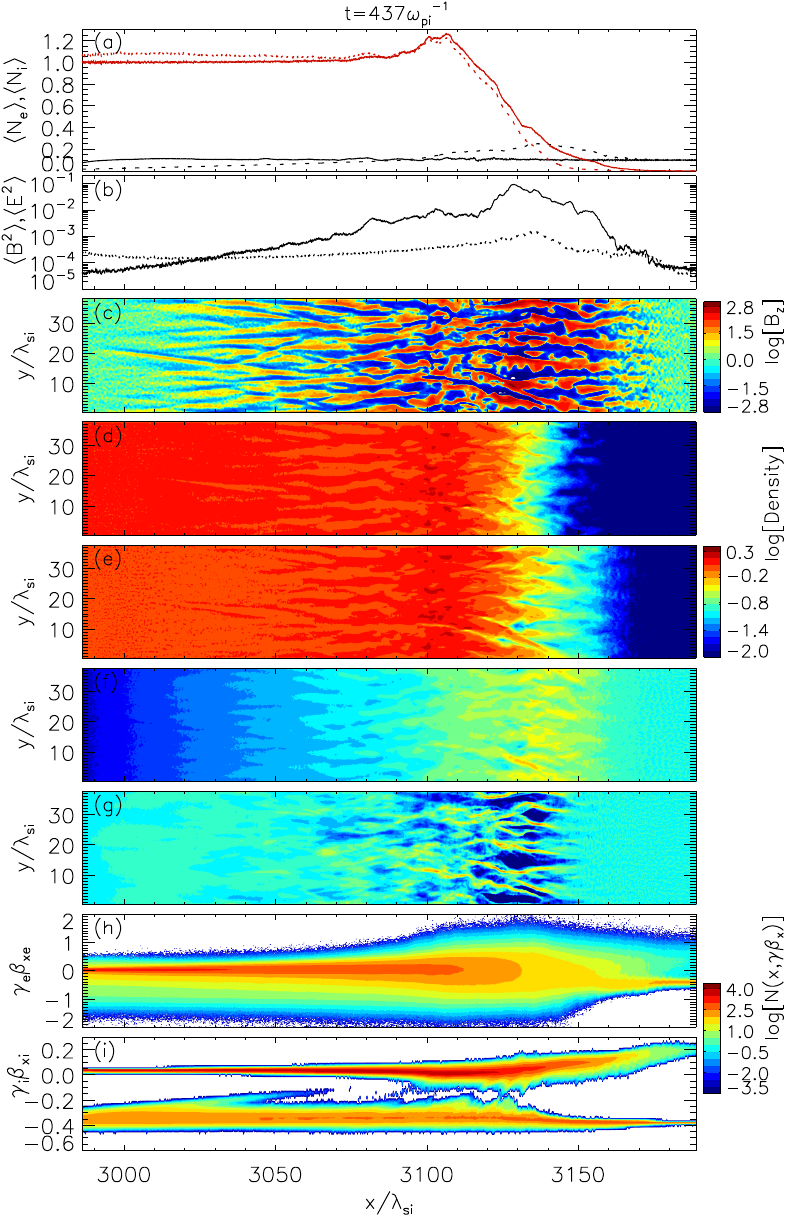}
\caption{Structure of the plasma-collision region at time  $t=437
\,\omega_{pi}^{-1}$
{for the unmagnetized run U1. Displayed are the}
profiles of (a) the average particle-number density normalized to {the
far-upstream} density of the dense plasma
(red lines: dense plasma, black lines: tenuous plasma; solid lines: ions, dotted
lines: electrons), (b) profiles of the average magnetic (solid line) and
electric (dotted line) energy density {in simulation units}, (c) the
amplitude of the magnetic field $B_z$ ({in sign-preserving logarithmic
scale as ${\rm sgn}(B_z)\,(3+\log\left[\max(10^{-3},\vert B_z\vert)\right])$)},
the density of dense electrons (d), dense ions (e), tenuous electrons (f), and
tenuous ions (g), all normalized to the {far-upstream} density of the dense
plasma, and the
longitudinal phase-space distribution of electrons (h) and ions (i).   
\label{unm1}}
\end{figure}

By the time $t\sim 400\,\omega_{pi}^{-1}$, the magnetic turbulence has reached
sufficient
strength that fresh incoming upstream electrons can no longer cross the
collision region. A contact 
discontinuity is formed that separates the electrons of both plasma slabs which
slowly accumulate 
at this barrier. The structure of the collision region at this early evolution
stage is shown in 
Figure \ref{unm1}. The CD is located around the initial collision boundary at
$x/\lambda_{si}\approx 3120$. 
As shown by the average-density profiles (Fig. \ref{unm1}a) and {particle}
density distributions 
(Figs. \ref{unm1}d-\ref{unm1}g), the CD creates a barrier for less energetic
dense ions. Only a few of these particles penetrate the tenuous-plasma region where they help
maintaining
charge neutrality. Since the number density of the dense ions {in the
tenuous-plasma region} is 
comparable to that of the {tenuous} ions, the former play an important
dynamical role in 
the system. The tenuous ions are only weakly perturbed upon crossing the CD. The
fast tenuous-ion 
beam propagates into the extended region to the left of the CD. Its net charge
is compensated in 
part by electrons of the dense beam that have been
reflected during the initial plasma encounter and in part by hot tenuous
electrons that leak
across the CD. Very few dense electrons cross the CD into the tenuous-plasma
region. In this 
zone, the charge balance with the dense ions streaming against the dilute beam
is provided by 
reflected tenuous electrons. 

The structure of magnetic turbulence shows different characteristics on the two
sides of the CD (Fig. \ref{unm1}c). In the dense-plasma region, the tenuous fast
beam constitutes a perturbation to the system. Ion-ion counterstreaming in this
{asymmetric} beam/plasma system renders the unmagnetized oblique instability
dominant in the spectrum of unstable modes. Oblique filamentary structures of
wavelength $\lambda\sim 3.5 \lambda_{si}$ can be seen in the transverse
component of the magnetic field and also through density modulations of the
electrons and ions {\citep{stroman}. 
The density filaments are stronger in the tenuous plasma because they can be
compensated by 
much weaker perturbations of the dense beam.}
Weak-beam conditions are also found at the leading edge of the dense ion beam
propagating into the tenuous-plasma region to the right of the CD. Here, the
dilute beam is composed of the most energetic ions of the dense slab that have
been accelerated in the collision region 
(Fig. \ref{unm1}i). The oblique mode is thus excited with wavelength 
$\lambda\sim 5 \lambda_{si}$, comparable to the skindepth of the tenuous-ion
component. However, the dense-ion beam shows a very strong density gradient in
the tenuous-plasma region. The conditions that determine the spectrum of
unstable modes thus vary with distance from the CD from {ones} suitable for
the oblique mode further upstream (at $x/\lambda_{si}\approx 3170$) to those in
the region closer to the CD (at $x/\lambda_{si}\approx 3150$) that allow
quasi-perpendicular filamentation{-like} modes {($k_{\parallel}\approx 0$)}
to grow fastest. 
In effect, the magnetic structure formed at the dilute head of the dense-ion
beam becomes strongly modified by a filamentation{-like 
instability which grows faster than the oblique instability and therefore}
leads to {more} efficient magnetic-field amplification. As seen in Figure
\ref{unm1}, the {instabilities spatially separate the ions of both slabs}.
Narrow filaments in cold tenuous ions are surrounded by electrons. The volume
between the filaments is depleted of tenuous ions and occupied by the ions of
the dense beam. The filamentary structures in 
the electron components are more diffuse on account of the {higher}
temperature of these particles. The turbulent magnetic structure advected toward
the CD is further amplified through mergers of current filaments. The observed
characteristics of the magnetic-field generation in the tenuous-plasma region
are thus reminiscent of the Weibel-mediated relativistic shocks in unmagnetized
plasmas {\citep[e.g.,][]{medv99,silva03,nishik03,fred04}}. 

Whereas the filamentation instability operating in the tenuous plasma can
generate strong magnetic turbulence in this region, 
the oblique instability leads to only moderate magnetic-field amplification in
the dense plasma left of the CD. As shown in Figure \ref{unm1}b, the energy
density of the magnetic field is more than an order of magnitude larger in the
region to the right of the CD, and this proportion is not significantly modified
during the subsequent evolution of the system. The strong magnetic field
generated in the CD region starts to efficiently deflect and isotropize
particles. The flow speed of the bulk plasma decreases, and larger numbers of
ions are reflected back toward their respective upstream regions (Fig.
\ref{unm1}i). The bulk kinetic energy of decelerating ions is converted into
electron heating (Fig. \ref{unm1}h). These processes of energy dissipation
should enforce the formation of a pair of shocks on the two sides of the CD. 

\subsubsection{Late-Stage Evolution of the Forward-Shock Transition}
\begin{figure*}[htb]
\epsscale{1.177}
\plotone{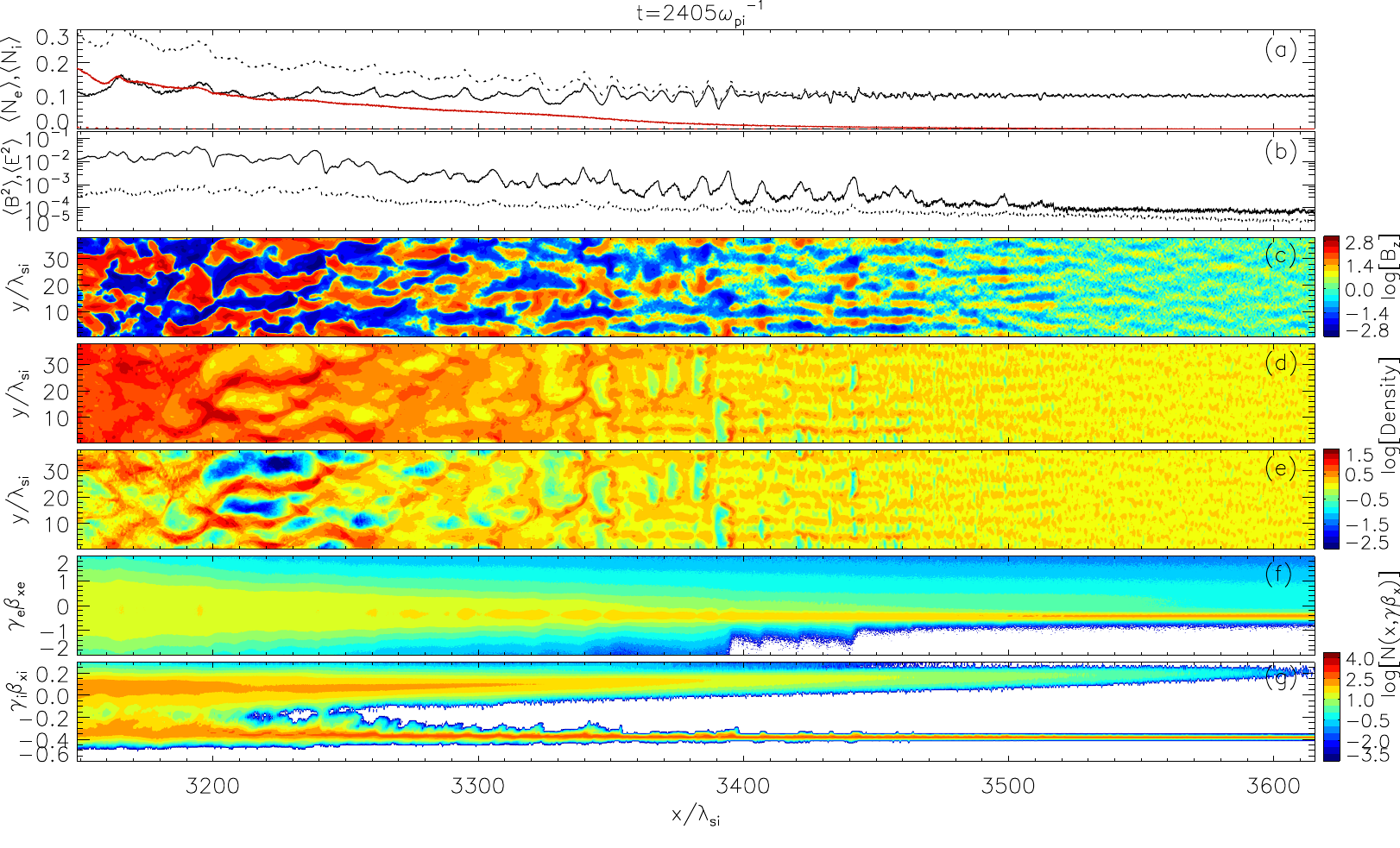}
\caption{Structure of the forward-shock precursor at time  $t=2405\,
\omega_{pi}^{-1}$.
{Shown are the}
profiles of (a) the average particle number density, (b) the average magnetic
and electric energy density (see Fig. \ref{unm1}a-b), (c) amplitude of
magnetic field $B_z$ {(logarithmic scale; see Fig. \ref{unm1}c)}, the
density of tenuous electrons (d), tenuous ions (e),
both normalized to the {far-upstream} density of the tenuous plasma, and the
longitudinal phase-space distribution of electrons (f) and ions (g).
\label{unm2}}
\end{figure*}

In the tenuous-plasma region, at $t\sim 500\,\omega_{pi}^{-1}$ the compression
ratio
reaches the level of $\sim 3$ predicted by the hydrodynamic jump conditions for
a two-dimensional nonrelativistic plasma with adiabatic index 
$\Gamma_{\rm U}=2$. The compression of electrons is provided solely by the
tenuous-electron component whereas the ion compression is initially dominated by
the dense ions (see Fig. \ref{unm2}). The extended region between the
unperturbed dilute plasma and the CD constitutes a forming forward shock. 

The structure of the forward-shock precursor (in the right half of the
simulation box)
is initially exclusively controlled by strongly nonlinear interactions between
counterstreaming ion beams resulting in the formation and mergers of current
filaments \citep[see, e.g.,][]{medv05,polomarov08,shvets09}. 
The corresponding turbulence is not purely magnetic and transverse.  
The two ion beams pinch at a different rate, on account of their different density and
temperature. Space charges {thus} arise that induce a weak
in-plane electric-field, $E_y$, which is not shown here
\citep{tzou06,fiore06}. {  In the nonlinear phase, the magnetic-pressure
gradient can also drive strong electrostatic fields \citep{2009PhPl...16g4502D}.}
The size of the magnetic precursor is thus determined by the range of the dense-ion
beam in the tenuous plasma. Ahead of the filaments, a population of hot
tenuous electrons, that have been reflected in the CD region, streams with
mildly-relativistic velocity against the incoming beam. This excites a
two-stream instability between reflected electrons and beam electrons that
produces mainly longitudinal (${\bf{k}\times \bf{v}_{\rm R}}=0$) perturbations
in $E_x$ 
{and associated electron charge-density modulations} 
with wavelength 
$\lambda \sim 5.5\, \lambda_{se,R}$, in agreement with our linear analysis.
\footnote{The linear analysis performed for conditions representative of the
far-upstream forward-shock precursor also shows the existence of a weak
filamentation mode that is in fact observed but does not provide a significant
dynamical impact on the precursor. The third predicted mode -- the Buneman mode
resulting from the drift of electrons against the ions, that is needed for
current balance with the reflected electrons, is an artefact of the
zero-temperature limit assumed in the calculations and should dissipate quickly
in the warm precursor plasma.} 
The amplitude of {electrostatic} perturbations saturates at small level and
their main effect is to heat-up the electrons {of the incoming beam} far
upstream of the forward shock.  
{However, deeper in the forward-shock precursor the electron-density
perturbations become nonlinearly enhanced through the approaching stream of the
dense ions, and the fluctuations in $E_x$ are amplified to
amplitudes far exceeding those further upstream. This leads 
to a strongly nonlinear backreaction on the tenuous-plasma ions, whose density
structure {begins to resemble} that of the electrons.}  
Figure \ref{unm2} presents the structure of the precursor region at time
$t=2405\,\omega_{pi}^{-1}$ that is representative of the characteristics
discussed above. The figure shows profiles of the average particle densities
(Fig. \ref{unm2}a) and electromagnetic energy densities (Fig.
\ref{unm2}b), 
a two-dimensional {distribution} of the magnetic field $B_z$ (Fig.
\ref{unm2}c), the densities of electrons and ions {of the tenuous plasma}
(Figs. \ref{unm2}d and \ref{unm2}e), and the electron and ion phase-space
density (Figs. \ref{unm2}f and \ref{unm2}g). The electrons of the dense plasma
are absent in this region. 
The density of the dense-ion beam is homogeneous and shows a strong gradient 
(see red solid line in Fig. \ref{unm2}a and the beam with positive momentum in
Fig. \ref{unm2}g). The far-upstream region is excluded {from Figure
\ref{unm2}}, but the hot tenuous electrons counterstreaming against the incoming
beam can be clearly seen as a population with positive momenta beyond
$x/\lambda_{si}\approx 3550$ in Figure \ref{unm2}f.
The {strong} density fluctuations in the ions and electrons of the tenuous
plasma can be seen in Figures \ref{unm2}d and \ref{unm2}e for $x/\lambda_{si}
\gtrsim 3450$. The longitudinal structures coexist with transverse 
density filaments indicating that the filamentation{-like} instability and
{the density modulation discussed {above} operate in parallel.} 
Closer to the shock, the density {modulations} {eventually lead}
to the formation of plasma cavities. {In this region,} the density of the
dense-ion beam is increased and also a population of reflected tenuous ions
exists
(for $x/\lambda_{si} \lesssim 3400$; the density of the shock-reflected ions is
more than a factor of 10 smaller than that of the dense ions).
{Such nonlinear structures are known to frequently occur in plasma physics and
astrophysics applications 
\citep[e.g.,][]{gold84}.}
As the parallel density structures are advected towards the shock and the
density of the dense-ion beam/shock-reflected ions increases, the modulations
grow further in amplitude and eventually begin to {rapidly expand radially}.
The expansion and subsequent merging of the adjacent cavities causes a
compression of the plasma between the voids and enhances electric currents in the
compression regions that lead to {additional}
magnetic-field amplification. 

The filamentary structure of the magnetic field thus becomes modified in the
shock precursor. 
{This can be observed}
in Figure \ref{unm2}c and in the mean magnetic-energy-density profile in Figure
\ref{unm2}b (solid line) in which the effects discussed are revealed as
localized magnetic-field amplification. Note, that although the interaction
between the dense-ion beam and the 
{charge-density waves}
turns highly nonlinear, it is the filamentation instability that remains the
dominant source of magnetic-field generation. As is evident in Figure
\ref{unm2}b, the average magnetic-energy density in the region of nonlinear
filamentation ($x/\lambda_{si} \lesssim 3280$) is about an order of magnitude
larger that the energy density associated with plasma compressions further
upstream. Nevertheless, the role of 
{density modulations} is not negligible. It is also important to
note that the volume filled with weak electrostatic turbulence ahead of the
magnetic-filamentation zone constitutes an inherent part of the shock precursor
and as such must be included in realistic PIC modeling. The electrostatic
precursor is not a transient effect in nonrelativistic shocks, because there
will always be a population of electrons reflected upstream {with}
relativistic velocities 
{that outrun the bulk of the shock-reflected ions.}

In our symmetric unmagnetized {test} run (equal density in the colliding
plasma slabs) we see similar effects in the shock precursor, but the cavities
represent weaker density fluctuations and do not grow nonlinearly as in 
{the asymmetric simulation U1}. Consequently, magnetic-field filaments are only
weakly modified {by plasma compression. The cavities that form upstream,
dissipate closer to the shock} {and give place} to density filaments.

\begin{figure}[htb]
\epsscale{1.178}
\plotone{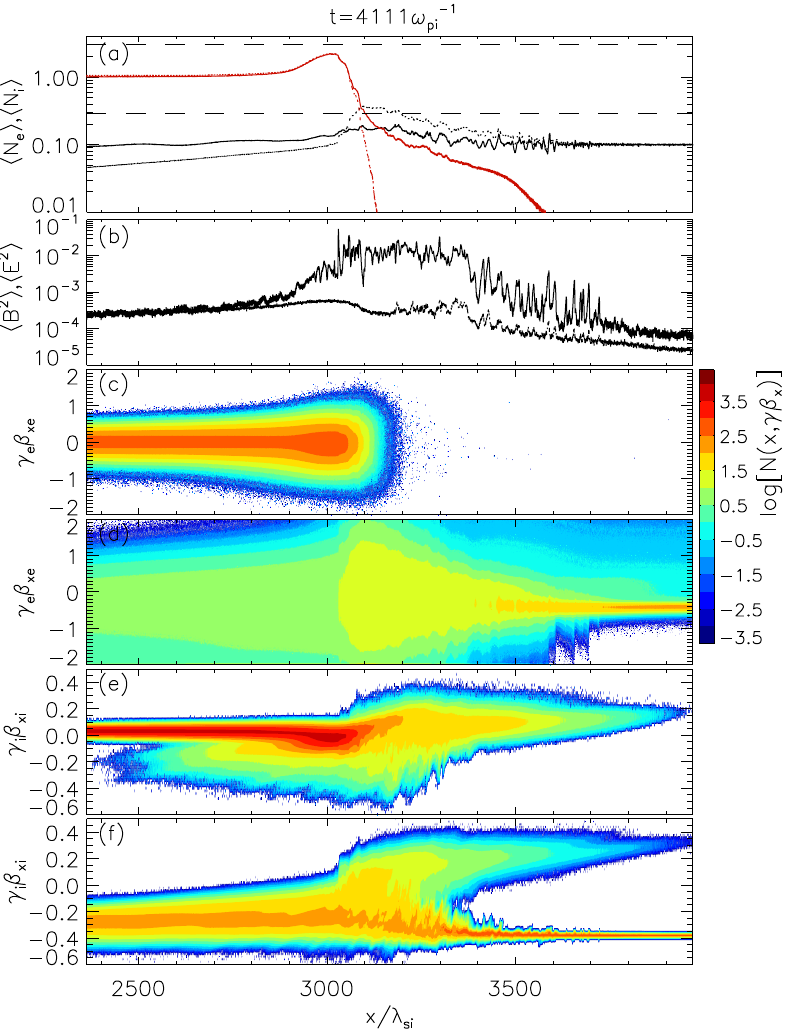}
\caption{Structure of the collision region at the end of the simulation U1 at
time  $t=4111 \,\omega_{pi}^{-1}$. {Shown are the}
profiles of (a) the average particle-number density {(a logarithmic scale is
used)}, and (b) the average magnetic and electric energy density (see Fig.
\ref{unm1}a-b), 
and the longitudinal phase-space distributions separately for all particle
species: electrons of the dense (c) and tenuous (d) plasma and ions of the dense
(e) and tenuous (f) component.
{In Fig. \ref{unm3}a, the horizontal dashed lines mark the hydrodynamic
compression level of
$N_{\rm fU,d}/N_{\rm fU,u}=2.9$ for the forward shock (lower line) and 
$N_{\rm rU,d}/N_{\rm rU,u}=3.01$ for the reverse shock (upper line).}
\label{unm3}}
\end{figure}

Figure \ref{unm3} shows the structure of the collision region at the end of the
simulation at time $t=4111 \,\omega_{pi}^{-1}$. The profiles of the average
particle densities (Fig. \ref{unm3}a) and electromagnetic-energy densities
(Fig. \ref{unm3}b) are presented
together with the phase-space distribution separately for all particle species
(Figs. \ref{unm3}c-\ref{unm3}f). 
The plasma density and electromagnetic-field structure (not shown) are analogous
to that at 
the earlier stage as described above (compare Figs. \ref{unm2}a-\ref{unm2}b with
 Figs. \ref{unm3}a-\ref{unm3}b). 
The CD is located at $x/\lambda_{si}\approx 3050$. By the end of the run, its
speed
approaches the predicted $\beta_{\rm CD}=-0.06$ in the simulation frame. The
forward shock region extends to right of the CD. The shock compression is in
agreement with a hydrodynamic value of
$N_{\rm fU,d}/N_{\rm fU,u}=2.9$ (lower dashed line in Fig. \ref{unm3}a), but
the shock front is poorly defined.

The plasma filaments that grow in size when approaching the forward shock should
disintegrate to form a nearly homogenuous downstream region. 
However, this stage is not yet observed by the end of the simulation. Although
the downstream tenuous-ion density distribution is strongly turbulent, it is
evident from Figure \ref{unm3}f that the dilute-ion beam is not decoupled from
the dense plasma and, despite being hot, continues to carry a substantial
bulk-flow energy across the CD. 
Consequently, the forward-shock compression ratio
carries a substantial contribution from the dense ions. 
 Note in Figure \ref{unm3}a, that the density of dense ions roughly equal to the
tenuous-ion density is required to match the electron compression, that is
provided solely by tenuous electrons.

\begin{figure}[htb]
\epsscale{1.178}
\plotone{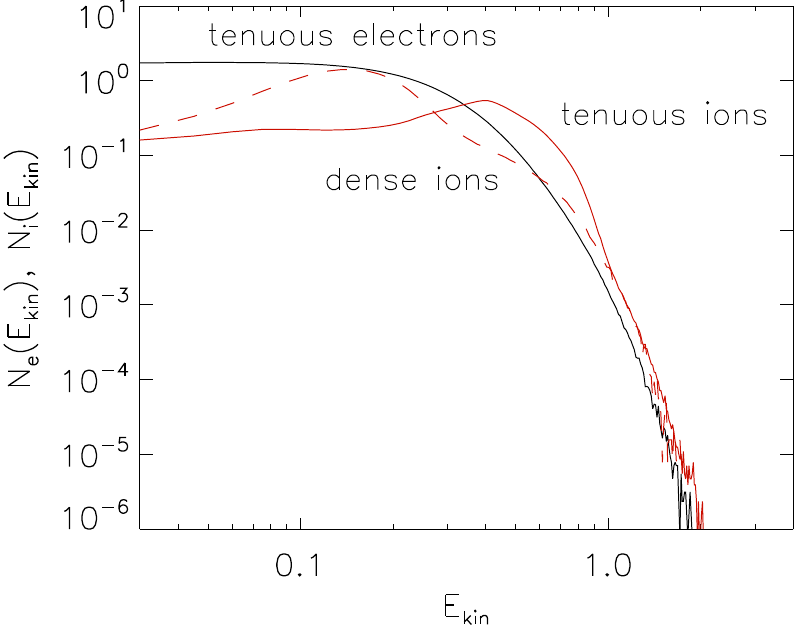}
\caption{Kinetic-energy spectra of particles downstream of the forward shock at
$x/\lambda_{si}\approx 3120$ and time $t=4111 \,\omega_{pi}^{-1}$ for the
unmagnetized run U1, in the CD rest frame. In simulation units $E_{\rm
kin}=0.25\,(\gamma-1)$ for electrons and $E_{\rm kin}=12.5\,(\gamma-1)$ for
ions. Normalized spectra are shown for tenuous electrons (black solid line),
tenuous ions (red solid line), and the dense ions (red dashed line).
\label{unm4}}
\end{figure}

 As shown in Figure \ref{unm4}, that displays particle distributions in the
forward shock downstream, the tenuous-ion spectrum strongly deviates from a
thermal distribution. 
The electrons of the tenuous plasma, accelerated at the expense of the ion
bulk-kinetic energy, are thermalized. Since they are forced to follow the ion
spatial distribution to maintain charge neutrality, their density is neither
uniform nor steady in the downstream region.
We conclude
that the forward shock has not yet fully developed. Although in asymmetric
unmagnetized-plasma collisions kinetic instabilities generate ample magnetic
turbulence in the 
tenuous-plasma region, the magnetic field is still too weak to efficiently
decouple the dilute-ion beam from the dense slab and to force {the ion-beam
reflection} needed to maintain the shock.  

The contribution of the reflected tenuous ions to the total density of ions
counterstreaming against the incoming tenuous beam and to the plasma compression
at the CD continuously increases during the simulation. However, downstream of
the forward shock the density of tenuous ions begins to exceed that of dense
ions only at a very late stage. Reflection of the tenuous ions is most effective
downstream and in the foreshock region with strong current filaments (see region
with $3100 \lesssim x/\lambda_{si} \lesssim 3300$ in Fig. \ref{unm3}), but takes
place also in the ion precursor further upstream where plasma cavities
nonlinearly grow and expand ($3300 \lesssim x/\lambda_{si} \lesssim 3700$). This
can also be observed in Figure \ref{unm2}g. One can note that the locations at
which the ions are decelerated and deflected well coincide with the density
compressions caused by the expanding cavities (compare, e.g.,
$x/\lambda_{si}\approx 3340$ and $3390$ in Figs. \ref{unm2}d-e 
and \ref{unm2}g). As the electron phase-space plots reveal (Figs. \ref{unm2}f
and \ref{unm3}d), the ion deceleration at these locations is accompanied by
localized electron heating. 

As noted above, in the simulation frame the system of forward and
reverse shocks separated by a CD moves toward the incoming dense-plasma beam
with velocity $\beta_{\rm CD}=-0.06$. This speed defines the \emph{CD frame} or
\emph{downstream rest frame}, in which particle spectra are calculated. 
Particle spectra in kinetic energy, $E_{\rm kin}=(\gamma-1)m_l\,c^2
\,{\rm(}m_l=m_e, m_i{\rm)}$, are presented in Figure~\ref{unm4}. They are based
on all particles located downstream of the forward shock in a slice of width
$\sim 20\, \lambda_{si}$, centered at $x/\lambda_{si}\approx 3120$. 
At this location, the contribution of dense electrons (not shown) to the total
number density and kinetic energy density is negligible. The distribution of
tenuous electrons is isotropic and quasi-thermal 
{with no evidence of a supra-thermal tail. Thus} the efficiency of any
electron-injection process that may have occurred in the forward-shock region in
the unmagnetized plasma is very small (compare Sec. \ref{mdist}).
The electrons account for 37\% of the total particle
kinetic energy, dense ions contribute another 38\%, and dilute ions carry the
remaining 25\%. The ion distributions are very anisotropic, which impacts the
energy spectra
(compare Figs. \ref{unm3}e-\ref{unm3}f).
In the spectrum of tenuous ions, the decelerated beam largely contributes to a
peak around 
$E_{\rm kin}\approx 0.4$; heated beam particles and particles returning upstream
build the high-energy tail. 
The dense ions form a separate colder distribution that is not mixed in
phase-space with the dilute component. The high-energy tail in the spectrum is
formed by particles that are reflected back to the dense plasma region. These
features again demonstrate that the forward shock is yet in the formation phase
and the processes of energy exchange/equilibration between the electron and ion
components are still ongoing.

\subsubsection{Reverse-Shock Transition}\label{urev}
An important feature to be noted from Figure \ref{unm3} is that the forward
shock is a region in which the ions of the dense beam are strongly deflected and
accelerated towards negative momenta, up to the initial momentum of the tenuous
ions (see Fig. \ref{unm4}). In consequence, these particles are able to cross
the CD back to the reverse-shock region to form there a population of hot and
energetic ions counterstreaming against the incoming dense-plasma beam. This
component of the dense ions can be clearly seen in the phase-space distribution
displayed in Figure \ref{unm3}e. The counterstreaming dense-ion beam is
generated during the entire evolution of the system, slowly building the plasma
compression associated with the reverse shock left of the CD. 

By the time the simulation ends, the reverse-shock transition has not yet
reached the compression ratio of a hydrodynamic shock (upper dashed line in Fig.
\ref{unm3}a).
Although the plasma density is homogeneous in the vicinity of the CD,
{particle distributions, including dense electrons,} are strongly
anisotropic. 
Upstream of the density compression, the oblique-mode instability amplifies the
transverse magnetic field resulting in structures analogous to those observed
earlier. The instability is generated in the entire extended region that is
populated with particles of the hot tenuous-ion beam. The magnetic field is very
weak and exerts no visible modulation to the plasma density. This is largely due
to temperature effects. The tenuous-ion beam is accompanied by a cloud of very
hot tenuous electrons that provide charge neutrality and also cause substantial
heating of the dense electrons.
The oblique-filamentary magnetic structures advected and compressed in the
reverse shock are further amplified through interactions with energetic
dense-ion particles reflected in the CD region, so that a weak magnetic
precursor to the reverse shock forms.

\subsection{Magnetized Plasma}\label{smag}
The strength of the mean magnetic field assumed for the magnetized run M1
has been chosen to 
characterize the physical effects introduced to the system described in Section
\ref{sumag} by the presence of the homogeneous flow-aligned field component. The
physical conditions represent a weakly-magnetized plasma, in which strong
shock waves should form. As checked with 
additional test runs, systems with lower magnetic-field strength show
characteristics  intermediate between those observed for run U1 and M1.

\subsubsection{Early Evolution and Filamentation-like Instability in the
Dense-Plasma Region}\label{emag}

\begin{figure}[htb]
\epsscale{1.178}
\plotone{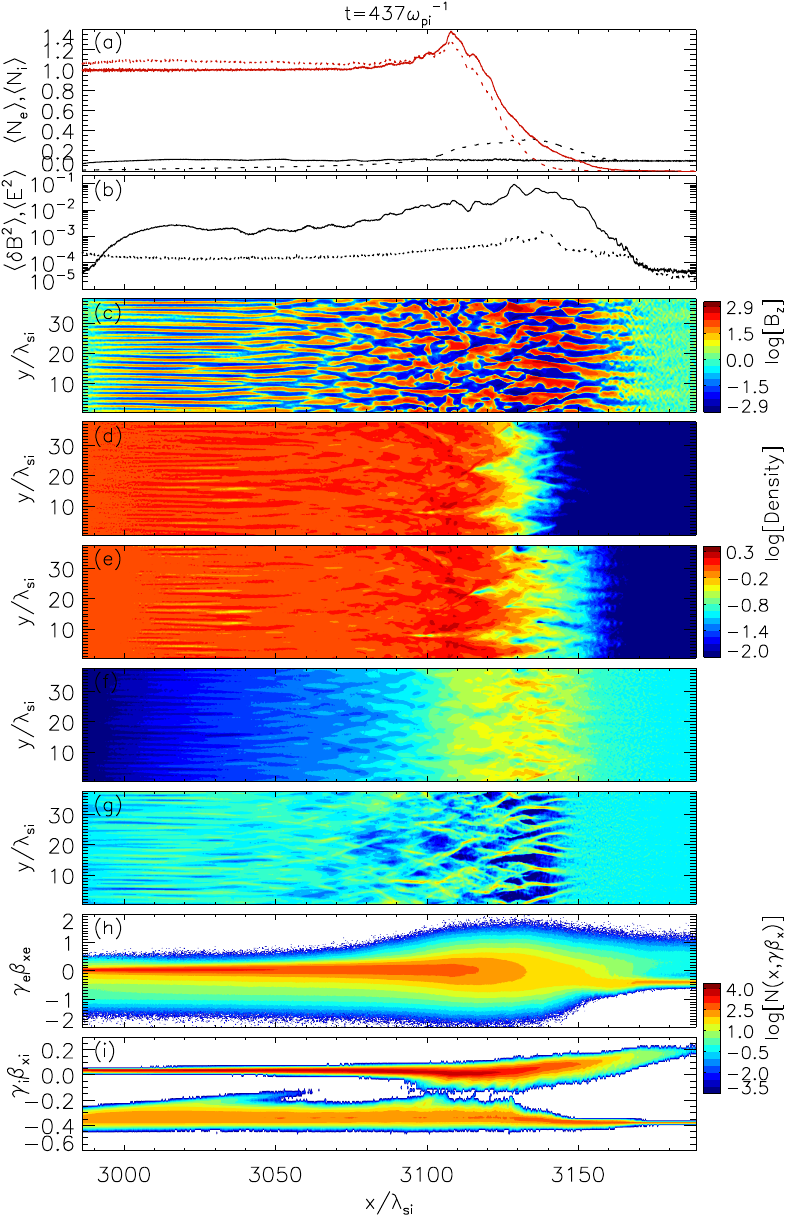}
\caption{Structure of the collision region at time  $t=437 \,\omega_{pi}^{-1}$
for the magnetized run M1 (see Fig. \ref{unm1}). Note that the mean magnetic
field is subtracted in the magnetic energy-density profile in Fig. \ref{mag1}b. The energy density associated with the mean field component is $7.9\times 10^{-3}$ in simulation units.
\label{mag1}}
\end{figure}

Figure \ref{mag1} shows the structure of the collision region in magnetized
plasma at early time, $t=437 \,\omega_{pi}^{-1}$. As in the unmagnetized
simulation (compare Fig. \ref{unm1}), the observed filamentation of plasma and
the associated perpendicular magnetic turbulence result from Weibel-like
instabilities. A direct comparison of Figures \ref{unm1} and \ref{mag1} reveals
that several characteristics of the magnetized system are analogous to those
described for the unmagnetized case. In particular, the magnetic-energy density
in the tenuous-plasma region exceeds the turbulent-field energy reached in the
dense plasma left of the CD. The similarities are also seen in the
electromagnetic and density structure
in the vicinity of the CD and in the tenuous-plasma region. 

Notably different from the unmagnetized case is the structure of the
dense-plasma region,
in which an almost purely perpendicular filamentation-like mode is produced at
the head of the 
tenuous-ion beam counterstreaming against the incoming dense plasma. This mode
appears in the 
out-of-plane magnetic field, $B_z$, that is strongly amplified. 
The large-scale distributions of particle densities (Figs.
\ref{mag1}d-\ref{mag1}g) show that 
the density modulations associated with this mode first appear in the tenuous
ions that form narrow filaments with large density contrast ($x/\lambda_{si}
\lesssim 3020$). Behind these structures, filamentary modulations develop in the
densities of dense electrons and ions ($3010 \lesssim x/\lambda_{si} \lesssim
3040$). These filaments merge and grow in size, but their density contrast
decreases further downstream of the tenuous-ion flow ($3040 \lesssim
x/\lambda_{si} \lesssim 3070$), where 
{magnetic-field filaments are further amplified through mergers.}

\begin{figure}[htb]
\epsscale{1.178}
\plotone{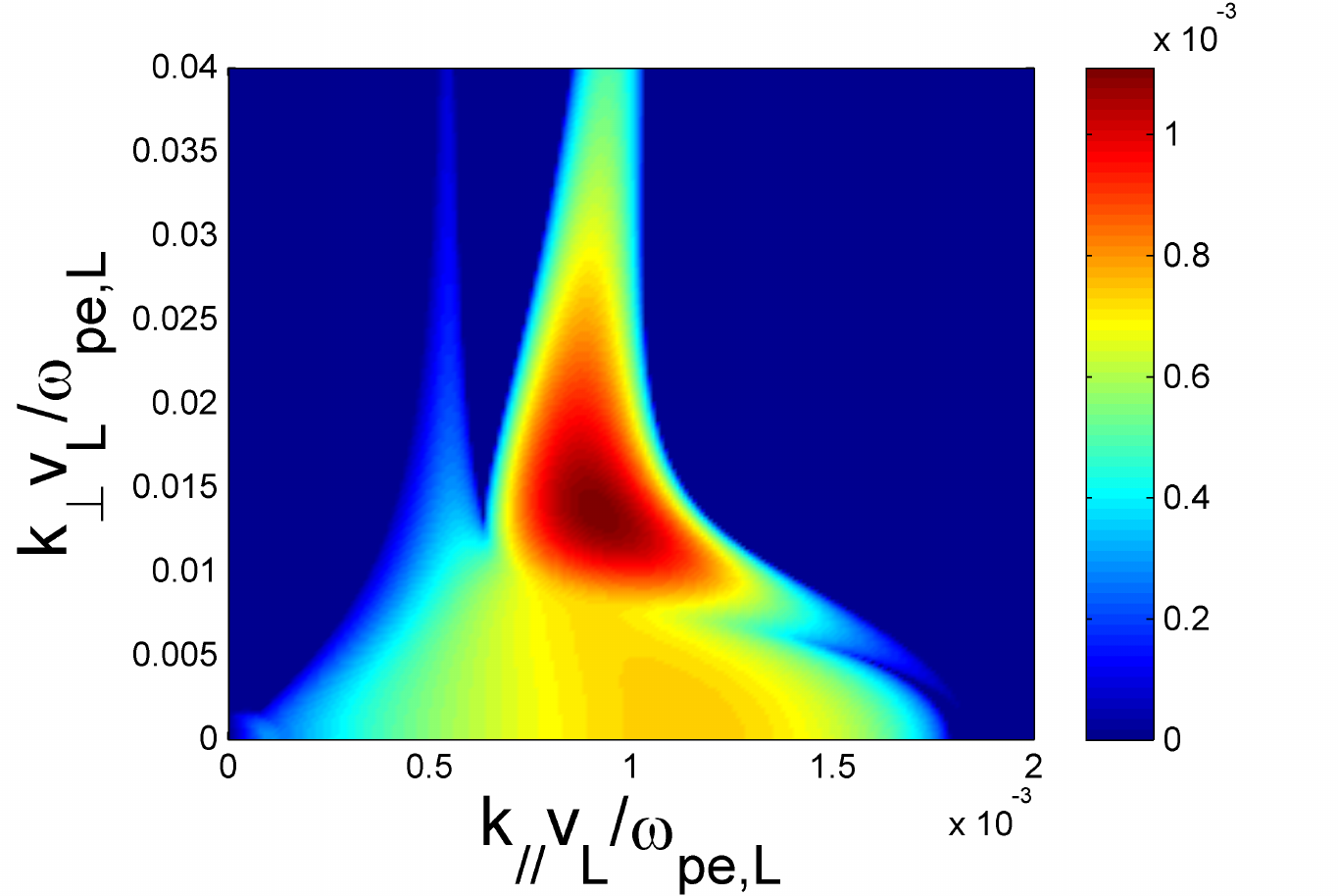}
\caption{Linear growth rate {(in units of the electron plasma frequency of
the dense plasma, $\omega_{pe,L}^{-1}$) of unstable modes {as function of the
flow-aligned and perpendicular wavenumber
($k_{\parallel}$ and $k_\perp$), assuming} parameters representing the
conditions at the head of the tenuous-ion beam propagating in the dense plasma
upstream of the reverse shock (run M1).} For a given $\mathbf{k}$, the growth
rate of the most unstable mode is plotted, {and we present only the low-$k$ part of the spectrum that does not contain}
two-stream electrostatic modes.
\label{mag1a}}
\end{figure}

The nature of the filamentation-like mode can be verified with linear analysis.
In Figure 
\ref{mag1a}, we show the growth rate in reduced-wavevector space $(Z_\parallel,
Z_\perp)$, $Z_i=k_iv_L/\omega_{pe,L}$, for parameters that should represent the
conditions at the head of the tenuous-ion beam. The calculations assume an
initial density contrast of $0.1$ between the tenuous and the dense-ion beam and
the corresponding initial injection velocities. The charge and current of the
dilute ion beam are to 90\% balanced by electrons of the dense beam reflected
during the initial plasma collision; the remaining portion is provided by
tenuous electrons leaking across the CD (see Fig. \ref{mag1}). The reflected
dense electrons are assumed to move with the bulk of the tenuous beam to account
for the current balance.
{Although the plasma beams are heated in the simulation, we perform our
calculations
in the zero-temperature approximation. 
However, in Figure \ref{mag1a} we 
{present only the low-$k$ part of the spectrum, related to instabilities that involve the
ions, and we do not show}
the two-stream electrostatic modes {at high wavenumbers}
that grow strongly in a cold plasma but are suppressed at finite temperature.}
The perpendicular wavevector of the dominant unstable mode, $Z_\perp = 0.0138$,
is in perfect agreement with $Z_\perp = 0.0125 - 0.0140$ measured in the
simulation ($\lambda\simeq 2.25-2.5 \lambda_{si}$) and so is a nearly purely
transverse character of the mode -- the predicted $Z_\perp/Z_\parallel\approx
15$. 
This suggests that the wavelength of the mode is not sensitive to temperature
effects, 
although the growth rate appears to be reduced in warm plasma.
{We have further confirmed this with additional test simulations that
assumed higher initial temperatures of the colliding beams (see also Sec.
\ref{revm}). 
Our large-scale run M2, with $m_i/m_e=20$, together with 
another set of test runs, that incorporated larger ion-electron mass ratios of
$m_i/m_e=100$ and 200, reveal that} a critical parameter determining the
strength of this unstable mode is ion inertia. 
{The effect of the magnetic turbulence in the collision region
on the motion of tenuous ions that pass through it is stronger for lower-mass
particles and results in a warm beam for run M2 (see also Appendix A). Instead,
for larger mass ratios the physical conditions in the test simulations
approximate a cold-plasma limit of the linear analysis.}

\begin{figure*}[htb]
\epsscale{1.177}
\plotone{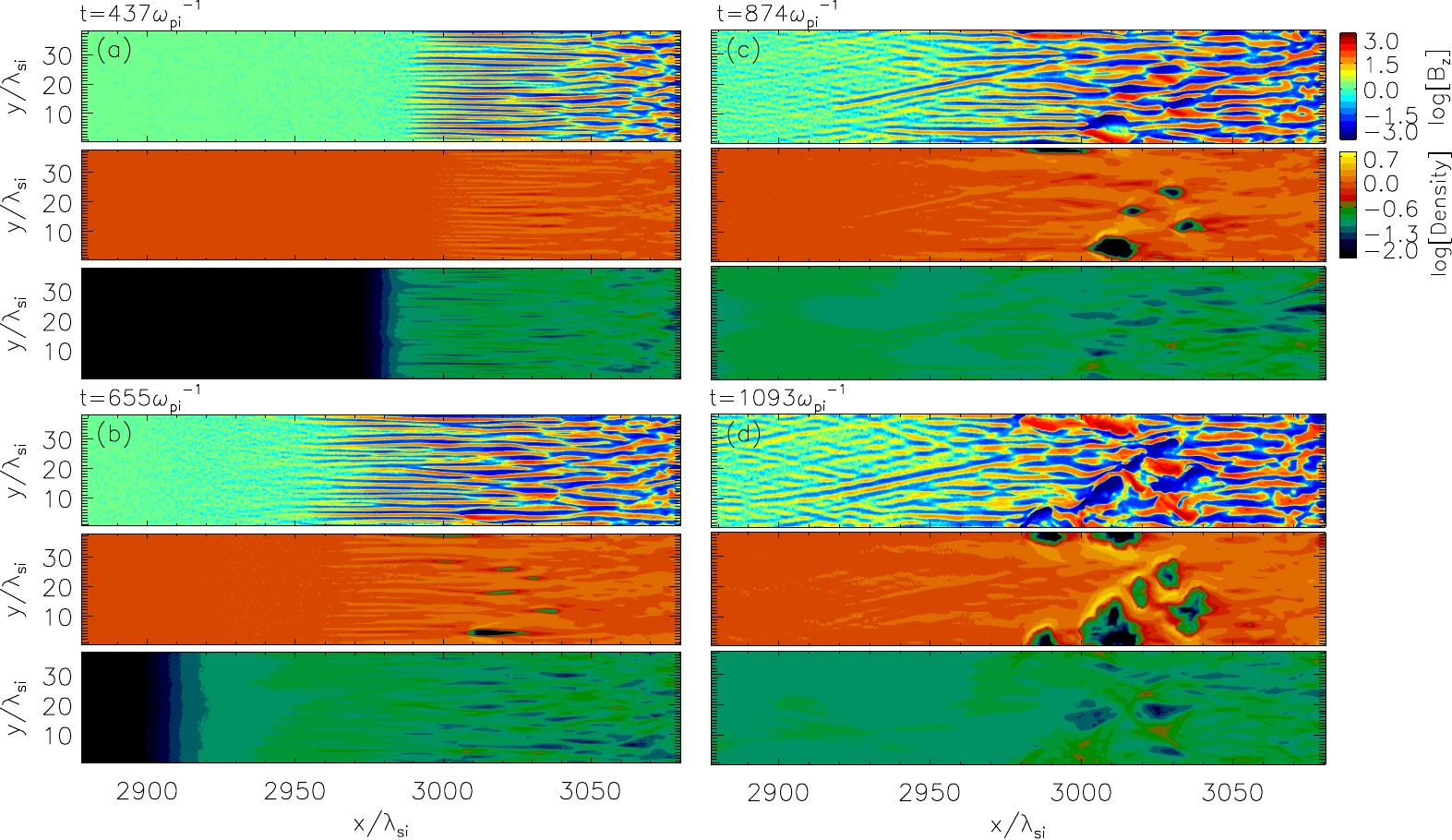}
\caption{Snapshots of the {nonlinear evolution of the filamentation-like
instability} 
in the dense-plasma region {upstream of the reverse shock at times (a)
$t=437$, (b) $655$, (c)
$874$, and (d) $1093\,\omega_{pi}^{-1}$; run M1.}  
From top to bottom, {each panel shows: the amplitude of the magnetic field
$B_z$ 
(in sign-preserving logarithmic scale; see Fig. \ref{unm1}c)}, the density of
dense ions, and the density of tenuous ions, {both normalized to the
far-upstream density of the dense plasma}.
\label{mag2}}
\end{figure*}

\begin{figure*}[htb]
\begin{center}
\includegraphics[scale=1.25]{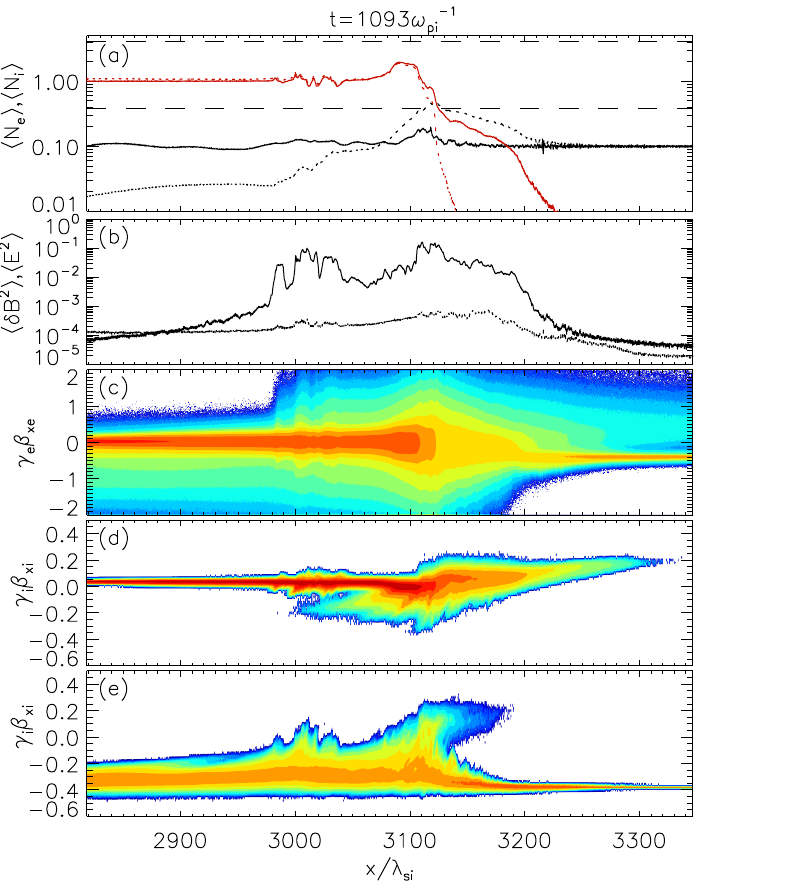}\hspace*{-2cm}
\includegraphics[scale=1.25]{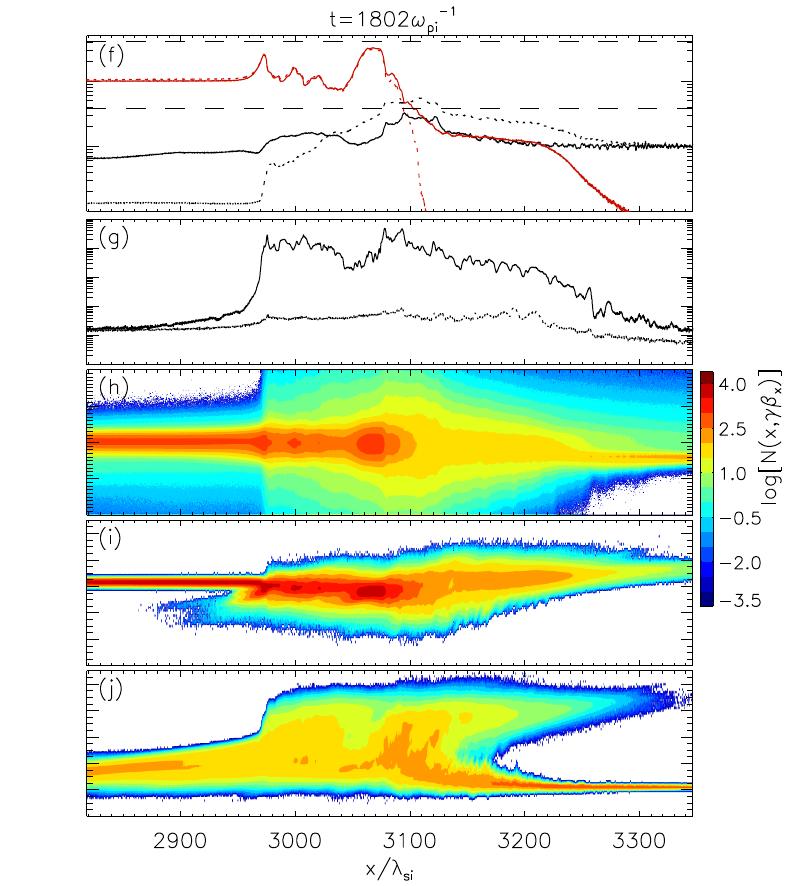}
\end{center}
\caption{Structure of the collision region {in the magnetized simulation M1,
at time $t=1093\, \omega_{pi}^{-1}$ (left panel; a-e; compare Fig. \ref{mag2}d)
and
$t=1801 \,\omega_{pi}^{-1}$ (right panel; f-j)}. 
{From top to bottom each panel shows:
profiles of the average particle-number density, profiles of the average
magnetic and electric energy density (see Fig. \ref{unm1}a-b), the
longitudinal phase-space distribution 
for both electron species, and the phase space separately for the dense and
tenuous plasma ions. 
In Figs. \ref{mag3}a and \ref{mag3}f, the horizontal dashed lines mark the
hydrodynamic compression level of
$N_{\rm fM,d}/N_{\rm fM,u}=3.86$ for the forward shock (lower line) and 
$N_{\rm rM,d}/N_{\rm rM,u}=4.02$ for the reverse shock (upper line).
The mean magnetic field with energy density of $7.9\times 10^{-3}$
is subtracted in the magnetic energy-density profiles in
Figs. \ref{mag3}b and \ref{mag3}g.}
\label{mag3}}
\end{figure*}

The presence of the filamentation-like mode significantly influences the
dynamical evolution of the system. The nonlinear development of the instability
in the dense-plasma region is presented in Figure \ref{mag2} as {a} sequence of
snapshots that show the $B_z$ magnetic-field component and density distributions
of dense and tenuous ions for $t\,\omega_{pi} = 437$, 655, 874, and 1093. The
distribution of dense electrons closely follows that of the dense ions. We also
show in Figure \ref{mag3} the average profiles of particle densities and
electromagnetic-energy {densities}, together with the phase-space
distributions across the whole collision region for 
$t\,\omega_{pi} =$1093 and 1801.
The filamentary modulations in the dense plasma, that were formed behind the
transverse tenuous-ion density filaments during the linear phase of the
instability ($t=437\,\omega_{pi}^{-1}$, Fig. \ref{mag2}a) turn in the later
stage into dense-plasma cavities surrounded by regions of enhanced plasma
density. These structures form only in the dense plasma and can be explained as
resulting from the magnetic bubbles that are created through localized
accumulation of magnetic energy caused 
by merging of
strong current filaments \citep[see, e.g.,][]{mur10}. The pressure gradient at
the magnetic bubbles pushes the dense plasma out and causes the growth of the
cavities. The subsequent radial expansion of the voids into each other results
in further plasma compression and thus magnetic-field amplification. 
Similar in appearance to the dense plasma voids, narrow filaments in the tenuous
ions, that {initially} emerge at the front of the beam, later merge, grow in size,
and eventually disperse
to develop a nearly homogeneous density distribution. 
The growth of the filaments can be traced through the magnetic-field structures
in Figure \ref{mag2}. {They demonstrate that the character of
the unstable mode evolves} into an oblique mode that
{provides} only a moderate magnetic-field amplification, compared to the
level acquired through the initial filamentation-like mode 
(compare the average magnetic-field profiles in Fig. \ref{mag1}b for
$x/\lambda_{si} \lesssim 3050$ and in Fig. \ref{mag3}b for $x/\lambda_{si}
\lesssim 2950$). 
The tenuous ions remain unaffected by the pressure gradient at the magnetic
bubbles due to their rapid motion. The density modulations in this plasma
component seen around 
the cavities 
are instead the result of the nonlinear development of streaming
instabilities that in this region become further enhanced by the approaching
denser stream of tenuous electrons and dense ions reflected in the vicinity of
the CD. This dense-ion component can be clearly seen in Figure \ref{mag3}d 
{for $3000 \lesssim x/\lambda_{si} \lesssim 3100$}.    

The formation of voids in the dense plasma greatly affects the subsequent
evolution of the system.
The cavities observed for $2970 \lesssim x/\lambda_{si} \lesssim 3050$ at
$t=1093\,\omega_{pi}^{-1}$ (Fig. \ref{mag2}d) can be seen as modulations in the
average density of the dense plasma, that are associated with a broad peak in
the magnetic-energy density (Fig. \ref{mag3}a-b) that is comparable to the
magnetic-turbulence level generated in the tenuous-plasma region to the right of
the CD. This strong field perturbs particle trajectories, thus enforcing the
reflection of the interpenetrating beams into their respective upstream regions.
One can observe this effect in Figures \ref{mag3}d-e. Localized particle
reflection in the region of the cavities is independent of beam deceleration and
reflection processes operating in the vicinity of the CD ($3050 \lesssim
x/\lambda_{si} \lesssim 3150$). Due to the large strength of the magnetic field,
which is continuously amplified by the compression between the expanding voids,
particle reflection is very efficient. One observes in the right panel of 
Figure \ref{mag3} that by the time $t=1801\,\omega_{pi}^{-1}$ returning ions of
both the dense and the tenuous plasma form almost homogeneous populations in the
extended region around the CD. The region of cavities thus constitutes a barrier
for the incoming beams that  
causes the accumulation of particles of each plasma component on their side of
this barrier. The average-density profiles (Fig. \ref{mag3}f) and phase-space
distributions (Fig. \ref{mag3}h-j) clearly illustrate this behaviour. Note in
particular the greatly increased contribution of the tenuous ions to the total
density of ions that counterstream against the incoming tenuous beam right from
the CD. Also, the dense-plasma region has become largely depleted of electrons
of the tenuous component, that in addition have experienced strong heating in
the region of cavities. 
  
The filamentation-like instability in the dense-plasma region thus provides a
mechanism to efficiently decouple the colliding ion beams. Compared to the case
of unmagnetized plasma, the presence of the mean magnetic field enables an
efficient formation of the forward and reverse shocks, whose subsequent
evolution is largely independent of the initial conditions.      
 
\subsubsection{Late-Stage Structure of the Double-Shock System}\label{dbl}

\begin{figure}[htb]
\epsscale{1.178}
\plotone{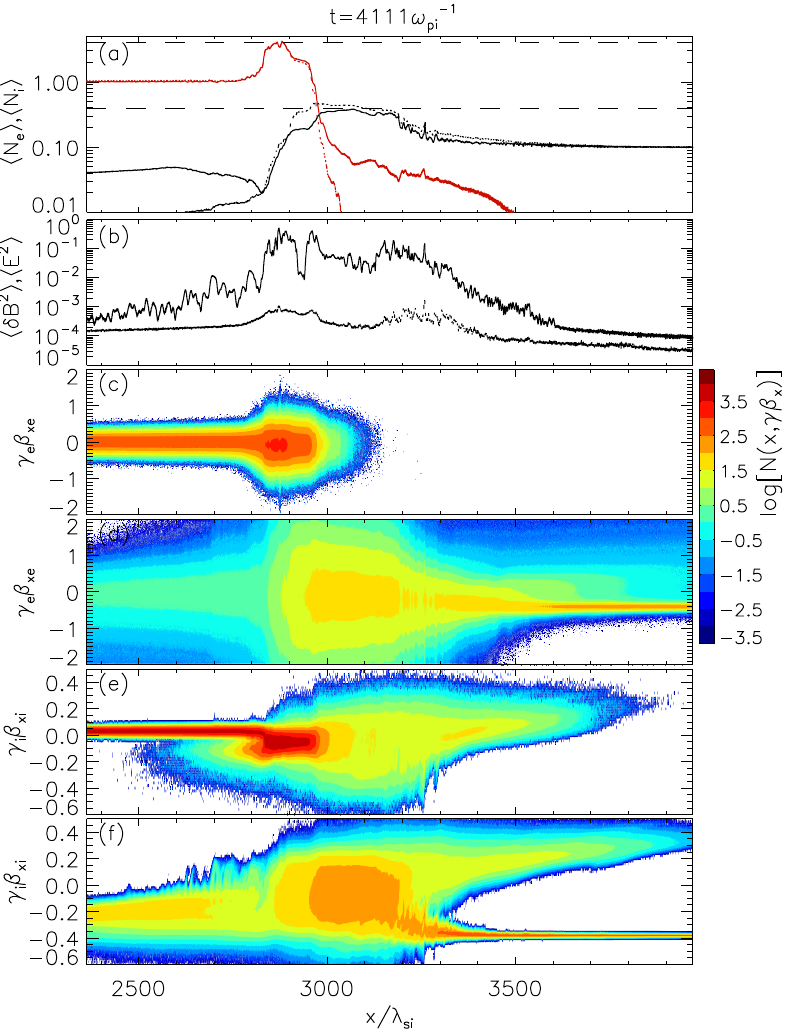}
\caption{Structure of the collision region at the end of the simulation M1 at
time  $t=4111 \,\omega_{pi}^{-1}$ {(see Fig. \ref{unm3}).
The horizontal dashed lines in Fig. \ref{mag4}a mark the hydrodynamic
compression level for the forward and reverse shock (see Fig. \ref{mag3}). The
mean magnetic field with energy density of $7.9\times 10^{-3}$ is subtracted in the magnetic energy-density profile in Fig.
\ref{mag4}b.}
\label{mag4}}
\end{figure}

The structure of the collision region at the end of the simulation at time
$t=4111\, \omega_{pi}^{-1}$ is presented in Figure \ref{mag4} (to be compared
with Fig.~\ref{unm3}). One can clearly see a system of forward and reverse
shocks separated by a CD. The shocks are visible as enhancement in the number
density at 
$x/\lambda_{si}\approx 3200$ and $x/\lambda_{si}\approx 2830$ for the forward
and reverse shock, respectively. The compression ratio at each shock reaches
$\sim 4$, in agreement with the hydrodynamical jump conditions for
nonrelativistic gas with $\Gamma_{\rm M}=5/3$, that predict
$N_{\rm fM,d}/N_{\rm fM,u}=3.86$ and $N_{\rm rM,d}/N_{\rm rM,u}=4.02$ 
in the simulation frame for the forward and reverse shock, respectively
{(upper and lower dashed lines in Fig. \ref{mag4}a)}.
The shock velocities measured in run M1 are also consistent with theoretical
steady-state 
values of 
$\beta_{\rm f,M}=0.04$ and $\beta_{\rm r,M}=-0.09$ in the simulation frame.  The
CD moves with a speed close to the predicted 
$\beta_{\rm CD}=-0.06$ and at the end of the run is located at 
$x/\lambda_{si}\approx 2950$. Note, that the location of the CD is shifted to
the left compared to that in the unmagnetized run. Only in the magnetized case,
the shock-formation processes allow for an efficient heating of the tenuous
plasma, whose pressure is sufficiently high to drive bulk-plasma motion
already at an early stage.
  
{As one can see in Figure \ref{mag4}a,}
the ion compression at the forward shock is almost entirely provided by the
tenuous plasma (cf. unmagnetized simulation U1). Although, as
evidenced by the ion phase-space distributions in 
Figures \ref{mag4}e and \ref{mag4}f, the contribution of the dense ions to the
number density of particles returning upstream is comparable to that of the
tenuous ions, the latter represent 
a more directed and faster flow deep in the forward-shock precursor and also
completely dominate the precursor dynamics further upstream (beyond
$x/\lambda_{si}\approx 3700$). 
Thus, late in the simulation the forward shock begins to resemble a
self-propagating structure that is exclusively maintained by shock-reflected
particles, independent of the processes operating at and behind the CD. This is
not yet the case in the dense-plasma region. Although the dynamics of the deep
reverse-shock precursor ($2750 \lesssim x/\lambda_{si}\lesssim 2820$) is solely
governed by the shock-reflected dense ions, the physics farther
upstream is still determined by the interaction of counterstreaming beams of
dense and tenuous ions (see below). Note, however, that the tenuous-ion beam in
the reverse-shock precursor, that has diluted, slowed-down, and heated, 
is at this stage decoupled from the main tenuous-plasma beam coming in from the
right of the CD (Fig. \ref{mag4}f). 

The turbulent electromagnetic field in the collision region is strongly
amplified throughout the simulation to eventually reach a level about an order
of magnitude larger than that achieved in unmagnetized conditions. One can see
in Figure \ref{mag4}b that the turbulence is almost purely magnetic, and both
the forward and the reverse shock develop strong magnetic precursors. The energy
density of the magnetic field in the forward-shock region peaks at the shock
front and then decays downstream \citep{kato08}. 
This turbulence is mostly generated when the forward shock is self-propagating.
The second maximum to the immediate right of the CD arises from the amplified
magnetic turbulence produced via filamentation instabilities during the early
evolution of the system, before the ion beams have become decoupled from each
other. The broad peak in the magnetic field energy density to the left of the CD
has largely evolved from the magnetic turbulence generated earlier in the region
of cavities (Fig. \ref{mag3}). It also constitutes the shock-compressed 
magnetic turbulence produced upstream of the reverse shock after the cavities in
the dense plasma have dissipated to form the downstream region of the reverse
shock.

\subsubsection{Structure of the Forward-Shock Transition}

\begin{figure*}[htb]
\epsscale{1.177}
\plotone{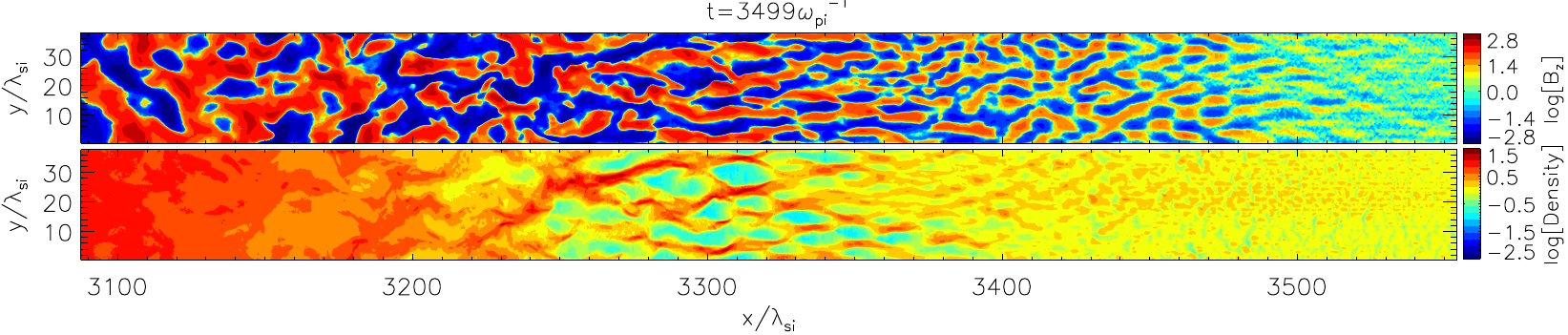}
\caption{Structure of the forward-shock precursor in run M1 at time  $t=3499\,
\omega_{pi}^{-1}$. {Shown is 
(a) the amplitude of the magnetic field $B_z$, (in sign-preserving logarithmic
scale; see Fig. \ref{unm1}c)}, and (b) the density of tenuous ions 
{normalized to the far-upstream density of the dilute plasma}.
\label{mag5}}
\end{figure*}

Figure \ref{mag5} presents the structure of the forward-shock transition at
time 
$t=3499\, \omega_{pi}^{-1}$. This time was chosen to display the {magnetic-field
amplitude}
$B_z$ (Fig. \ref{mag5}a) and the density of tenuous ions (Fig. \ref{mag5}b) with
correct aspect ratio; the characteristics discussed are {the same as} those
observed at $t=4111\, \omega_{pi}^{-1}$ (Fig. \ref{mag4}). The tenuous-electron
distribution (not shown) closely follows that of tenuous ions; they also provide
charge balance to the low-density dense ions that occupy the volume between
tenuous-ion filaments. As seen in Figure \ref{mag5}, the main source of
turbulent magnetic field is the electromagnetic Weibel-type instability.
Strong nonlinear effects of the interaction of returning particle beams with 
electrostatically-induced parallel density structures, reported in the
forward-shock precursor for the unmagnetized case, are not observed. The
two-stream instability between incoming and returning electron beams
generates electrostatic perturbations far upstream 
{that lead to nonlinear charge-density waves}
in the precursor (for $x/\lambda_{si} \gtrsim 3450$ in Fig. \ref{mag5}b), but
the interaction {of the density modulations with the reflected ions 
does not result in the formation of plasma cavities}.
The growth and expansion of {parallel} density fluctuations is
suppressed by the mean magnetic field.  

The tenuous-plasma filaments, that pinch and merge in the forward-shock foot,
break up at the
shock interface and dissipate to form a nearly homogeneous density distribution
downstream 
of the shock ($x/\lambda_{si}\lesssim 3150$ in Fig. \ref{mag5}b). The
disintegration of the tenuous-ion beam downstream of the forward shock was not
observed in the unmagnetized simulation U1. 
Nevertheless, although the downstream tenuous-ion density is close to uniform,
they are not fully thermalized (see below), which indicates that the
forward-shock evolution has not fully reached a steady state by the end of the
simulation.   

\subsubsection{Structure of the Reverse-Shock Transition}\label{revm}
As discussed in Section \ref{emag}, strong fluctuations  in the density of the
dense plasma and the magnetic field 
resulting from a filamentation-like instability in the region of cavities began
to form an
obstacle for the incoming tenuous plasma and eventually caused a depletion of
these particles in the dense-beam region upstream of the dense-plasma voids. In
addition, the tenuous ions at the head of the dilute-ion beam were heated and
the filaments dispersed. In effect, the character of the unstable magnetic mode
has changed from almost purely perpendicular to slowly-growing oblique 
turbulence, with characteristics similar to those described for the
reverse-shock precursor in the unmagnetized run U1 (Sec. \ref{urev}). As in the
unmagnetized case, 
the oblique magnetic mode is generated wherever the tenuous-ion beam propagates
in the upstream region of the reverse shock. However, the presence of the mean
magnetic field triggers an additional plasma mode in the system that slowly
grows and eventually dominates the magnetic-turbulence spectrum in the
reverse-shock precursor.  

\begin{figure}[htb]
\epsscale{1.178}
\plotone{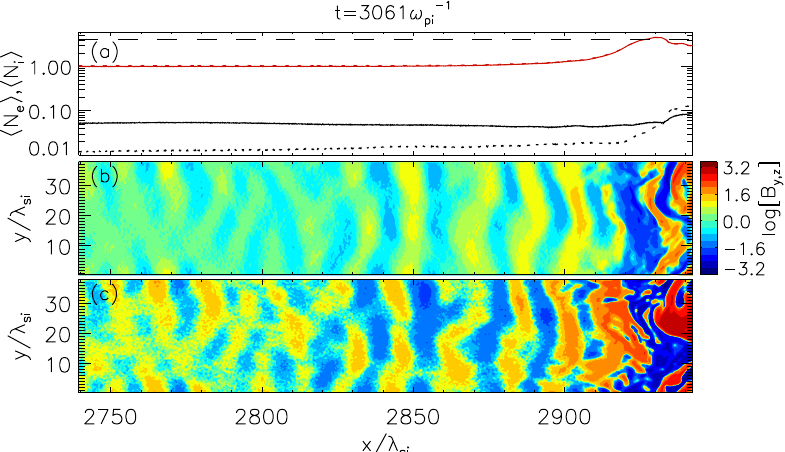}
\caption{{Structure of the reverse-shock transition in run M1 at time 
$t=3061\, \omega_{pi}^{-1}$. Shown are (a)  
profiles of the average particle-number density (see Fig. \ref{unm1}a) and the
amplitude of magnetic-field components, $B_y$ (b), and $B_z$ (c). 
The horizontal dashed line in Fig. \ref{mag6}a marks the hydrodynamic
compression level 
of $N_{\rm rM,d}/N_{\rm rM,u}=4.02$ for the reverse shock.
Note that the field
components are in sign-preserving logarithmic units, e.g., for $B_z$ as 
${\rm sgn}(B_z)\,(3+\log\left[ \max(10^{-3},\vert B_z\vert)\right])$.}
\label{mag6}}
\end{figure}

Figure \ref{mag6} presents the reverse-shock transition at time 
$t=3061 \,\omega_{pi}^{-1}$, at which the initial oblique mode has already
attenuated.
As seen in Figures \ref{mag6}b and \ref{mag6}c, that show the $B_y$ and $B_z$
magnetic-field components, respectively, the unstable mode represents an almost
purely parallel magnetic wave.
The wave is left-hand circularly polarized%
\footnote{For ${\bf{B}_{\rm 0}}=B_{\rm 0,x}{\bf{\hat{x}}}$ and
${\bf{k}}=k_x\,{\bf{\hat{x}}}$, a plane electromagnetic wave of the form
${\bf{B}}_1\exp i({\bf{k}}\cdot{\bf{x}}-\omega t)$, 
${\bf{B}}_1= B_{\rm 1,y}\,{\bf{\hat{y}}}+B_{\rm
1,z}\,{\bf{\hat{z}}}$
is left-hand circularly polarized if $B_{\rm 1,z}=-i\,B_{\rm
1,y}$.} 
and in the rest frame of the dense-plasma beam moves 
antiparallel to the mean magnetic field with phase velocity 
$v_{\rm ph}\approx 7.9\times 10^{-3}\,c\simeq 1.4\, v_{\rm A,L}$. Therefore, in
the frame of the reverse shock the wave is carried towards the shock with the
plasma flow. The wavelength of the mode is
$\lambda\sim 15\,\lambda_{si}$ and its frequency {is} $\omega\simeq 0.58\,\Omega_i$. 
These characteristics do
not significantly vary over the precursor region.

\begin{figure}[htb]
\epsscale{1.178}
\plotone{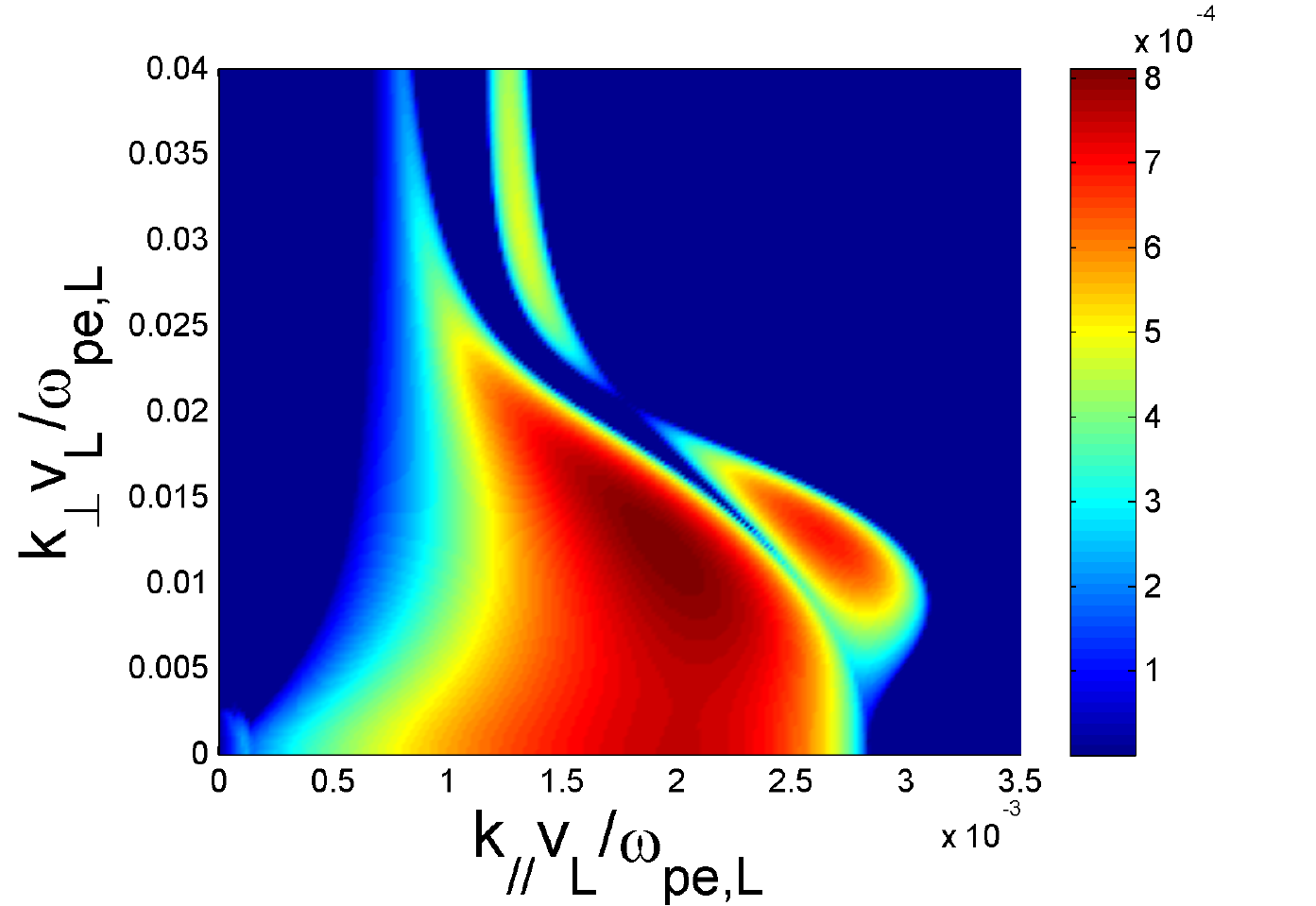}
\caption{{Linear growth rate of unstable wave modes
for conditions representing the late-stage structure of the reverse-shock
precursor in run M1.
(see Fig. \ref{mag1a})}.    
\label{mag6a}}
\end{figure}

To understand the nature of the observed parallel mode, a linear analysis has
been performed. 
{It is analogous to the analysis for the filamentation-like mode in Section
\ref{emag}, 
but uses a modified set of parameters, for the physical conditions in the
reverse-shock precursor have changed. The calculations thus assume {a} density
contrast of 0.05 between the dilute and the dense-ion beam (compare the average
density profiles in Fig. \ref{mag6}a) and {a} tenuous-ion beam velocity of
$v=-0.3\,c$,} slightly slower than the initial injection velocity $v_R$.
The charge and current of the tenuous ions is compensated mainly by
returning dense electrons;
electrons of the tenuous plasma represent only 5\% of all returning electrons. 
We assume the cold-plasma approximation. The results are displayed in Figure
\ref{mag6a}, in which the two-stream electrostatic modes {at high wavenumbers} are again
{not shown}.

The area of strongest growth at $Z_\parallel\approx 0.0019$ and $Z_\perp\approx
0.012$ pertains to 
an oblique mode, {analogous to the nearly transverse mode
discussed in Section \ref{emag}.} 
The second maximum 
at $Z_\perp\approx 0$ and $Z_\parallel\approx 0.002$, {or}
$\lambda_\parallel\approx 16\,\lambda_{si}$, can be identified with our parallel
mode. 
{Note, that the growth rates of both the oblique and the parallel modes are
comparable 
to within 6\% in the present setup,
which was not the case for the analysis of the early-stage turbulence modes in
the dense-plasma
region, in which a parallel mode can also be observed at $Z_\parallel\approx
0.0012$
but with {a} growth rate 70\% of that of the dominant unstable
filamentation-like mode
(Fig. \ref{mag1a}).}

{The nature of the parallel mode can be explained by analysis of the
dielectric tensor $\mathbf{T}(\mathbf{k},\omega)$ for our
system.} 
For flow-aligned perturbations with $k_y=0$, the tensor $\mathbf{T}$ takes the
form,
\begin{equation}\label{tensor}
\mathbf{T}= \left(
\begin{array}{ccc}
 b  &  0   &  -c \\
 0  &  a   &  0 \\
 c  &  0   &  b
\end{array}
\right),
\end{equation}
where the tensor elements $a,b,c$ are reported in Appendix \ref{ap:tensor}. As
one can see, the dispersion equation displays three branches, as $\det
\mathbf{T}(\mathbf{k},\omega) =0$ readily gives
\begin{equation}
a(b+ic)(b-ic)=0.
\end{equation}
The equation $a=0$ defines two-stream like electronic modes at large
wavevectors. {As already noted}, at the time-scale of the present
simulations, these modes quickly saturate {for warm electron beams}. On the
ion time-scale, the large-wavelength modes governing the system arise from the
branches  $(b+ic)=0$ and $(b-ic)=0$.
When analyzing a simpler setup formed by a dilute-{ion} beam passing through
an electron-ion plasma, the so-called nonresonant (Bell) mode, as well the
resonant one, both come from 
$(b-ic)=0$ \citep{Bell2004,Reville2006,winske}. In the present case, the
growth-rate profile observed along the parallel axis in Figure \ref{mag6a}{
(and also in Fig. \ref{mag1a})} can equally be traced back to the same
$(b-ic)=0$ branch. We can 
{therefore identify our parallel mode with} the "Bell-like" mode.
However, {the linear analysis predicts a frequency
$\omega=0.0012\,\omega_{pe,L}=0.0085\,\omega_{pi,L}\simeq 1.5\,\Omega_i$, that} is a
factor of $2.6$ higher than observed in our simulation. This indicates that
either the cold plasma approach is {invalid} or the simplified conditions
assumed for the linear analysis do not properly represent the conditions in the
simulations.

To evaluate how well the simplified conditions correspond to the state of the
plasma system in the simulated reverse-shock precursor and to estimate the
effects of a finite temperature, two additional test runs have been performed.
Both assume the conditions used in the
linear analysis and periodic boundary conditions. In the first run all plasma
components are cold. We refer to this simulation as to the {\it cold test-run}.
In the second test simulation, the plasma beams initially have temperatures as
measured in the precursor. This run is referred to as the {\it warm test-run}.
We found that both test runs well reproduce the characteristics of the initial
oblique mode, which is the fastest-growing mode in the linear analysis. The
nearly homogeneous density distributions, and the absence of strong density
filaments, that are observed in the main simulation, are both well replicated in
the warm test-run. In the latter and in the magnetized run M1, the oblique mode
is observed mainly through filamentation in the $B_z$ field component. 
The parallel mode appears at a later stage in both test runs. Whereas in the
warm run it is the only magnetic mode, in the cold test-run the oblique-mode
nonlinear filaments in  $B_z$ are clearly imprinted onto
the parallel magnetic-field fluctuations. These characteristics are due to
thermal effects that lead to damping of the wave modes with perpendicular
components, because the plasma pressure opposes pinching of the filaments.
Consequently, the oblique mode eventually dissipates in the warm plasma, and the
parallel mode, whose linear growth rate is only slightly lower than that of the
oblique mode, dominates the nonlinear evolution.

The parallel mode observed in the test runs has a wavelength
$\lambda_\parallel\sim 16\,\lambda_{si}$ ($Z_\parallel\simeq 0.002$), is
left-hand circularly polarized, and propagates 
antiparallel to the mean magnetic field. The measured growth rate is about half
the analytically-derived growth rate, which is consistent with the influence of
thermal effects. These results very well reproduce the properties of the
parallel mode observed in the
magnetized simulation. 
The measured frequency of the mode is $\omega\approx 1.1\,\Omega_i$ in the cold
test run and
$\omega\approx 0.7\,\Omega_i$ in the warm test-run. The results of the warm test
run are therefore consistent with our main simulation, which indicates that the
simplified conditions well represent the real conditions in the simulations and
thus our identification of the parallel mode {with the Bell-like mode} is
correct. The differences between the measured mode 
frequency and that predicted in the linear analysis likely result from thermal
effects, that in the nonlinear stage modify the physical conditions, possibly
introducing factors not accounted for in the {linear calculations}.

As one can note in Figure \ref{mag6}, the amplitude of the parallel
magnetic-field fluctuations considerably increases in the reverse-shock foot.
The latter is the precursor region in which a high-density population of
shock-reflected dense ions is present at
$x \gtrsim 2880\,\lambda_{si}$. The amplification and compression of the
parallel fluctuations
thus provides a formation mechanism for the reverse-shock transition that is
consistent with the hydrodynamic picture. Note, that although the parallel
magnetic mode excited in the precursor region results from the initial
conditions, similar unstable modes should appear later on account
{of shock-reflected dense plasma. In fact, our linear calculations show that
the appearance of
the Bell-like mode {critically} depends on the presence of \emph{returning
dense electrons}
in the shock-precursor plasma, and not on the tenuous-ion beam.}  
Therefore, one may expect that {the} further evolution of the reverse-shock system
will be similar.

\subsubsection{Particle Distributions and Electron Injection at the Forward
Shock}\label{mdist}
As in the case of unmagnetized plasma (run U1), particle
distributions are calculated in the 
rest frame of the downstream plasma, defined by the CD speed $\beta_{\rm
CD}=-0.06$ 
in the simulation frame.
Although in run M1 the downstream plasma is turbulent on large scales in the CD
rest
frame, these turbulent motions are slow and do not significantly distort the
spectra.

For run M1 we define the forward-shock downstream region as the area of roughly
constant plasma compression with ratio $\sim 4$, that is dominated by particles
of the tenuous plasma whose bulk velocity is similar to that of the CD. 
At $t=4111\, \omega_{pi}^{-1}$, the downstream region spans the range 
$2980 \lesssim x/\lambda_{si} \lesssim 3080$ (see Fig. \ref{mag4}). 
Left of this location is the CD region in which both the dense and the tenuous
plasma reside and where additional processes may operate that alter the particle
spectra.
As for run U1, to calculate particle distributions we choose a slice of width
$\sim 20\, \lambda_{si}$, now centered at $x/\lambda_{si}\approx 3000$. 
The slice is stationary with respect to the CD, and thus the shock front moves
steadily away from the location at which spectra are measured.

\begin{figure}[htb]
\epsscale{1.178}
\plotone{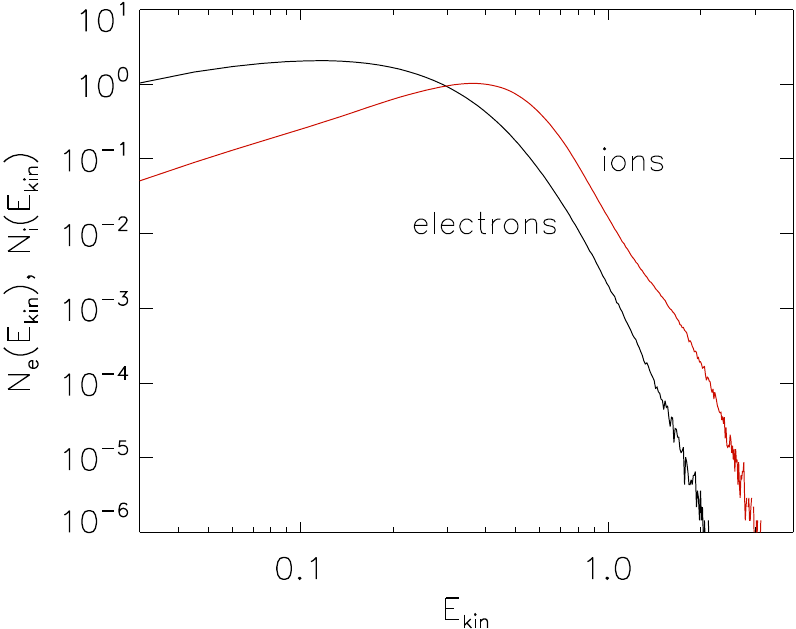}
\caption{Kinetic-energy spectra of tenuous electrons (black line) and tenuous
ions (red line) in the forward-shock downstream region at $x/\lambda_{si}\approx
3000$ and time $t=4111 \,\omega_{pi}^{-1}$ for the magnetized run M1, in the CD
rest frame. The spectra are normalized and expressed in simulation units, in
which $m_e\,c^2=0.25$.   
\label{mag7}}
\end{figure}

Figure \ref{mag7} presents spectra of tenuous electrons and ions at time 
$t=4111\, \omega_{pi}^{-1}$. Only the distributions of electrons appear
approximately thermalized. 
Visible also in Figure~\ref{mag4}, the distribution function of tenuous ions
exhibits a plateau when integrated over any two momentum coordinates, e.g.,
{$f(p_x)=\int dp_y\,dp_z\ f(\vec p)$,} and consequently the kinetic-energy spectrum
strongly deviates from a thermal distribution. Isotropy is approximately
maintained in the plateau section of the distribution, but not in the wings
where the distribution function falls off. The form of the distribution suggests
that the ions have reached stability against parallel electrostatic and
electromagnetic instabilities.
As in the case of the unmagnetized system, in run M1 the total energy content in
electrons is roughly the same as that in ions, in fact 2/3 of it. 

\begin{figure}[htb]
\epsscale{1.178}
\plotone{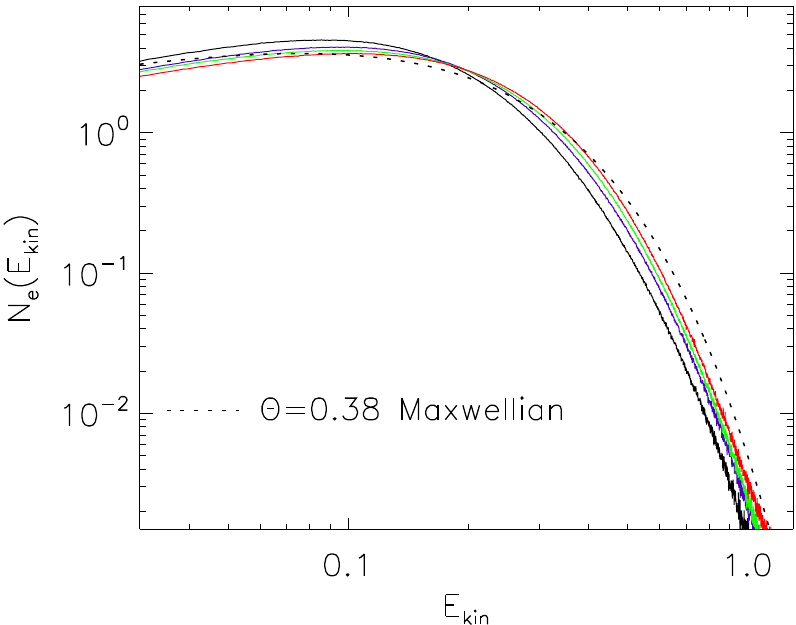}
\caption{Tenuous-electron kinetic-energy spectra in the forward-shock downstream
region at $\sim 50\lambda_{si}$ from the CD at times $t=2406$ (black line), 2974
(blue line), 3542 (green line), and  $4111\, \omega_{pi}^{-1}$ (red line) for
run M1. All spectra are normalized and expressed in simulation units, in which
$m_e\,c^2=0.25$. The dotted line shows a relativistic Maxwellian for comparison.
\label{mag8}}
\end{figure}

Figure \ref{mag8} details the time evolution of the downstream electron
spectrum. 
At all times for which spectra are presented ($t\,\omega_{pi}=2406$, $2974$, $3542$,
and
$4111$, shown with black, blue, green, and red line, respectively), the forward-shock
compression is well established in the extended region to the right from the CD,
and the electrons are isotropic in the downstream frame (cf. run U1). These
spectra display a slow increase in average energy, possibly because the region,
in which
the distribution is calculated, is located closer to the shock front at earlier
times.
All spectra are slightly more peaked than a relativistic Maxwellian, which we
plot for comparison. It is interesting to note that renormalizing the spectra to
dimensionless energy $x=E/E_{\rm kin, av}$, 
{$E_{\rm kin, av}$ being the {average particle} energy}, 
indicates that the shape of the
spectra evolves very weakly, and it is indeed mainly a continuous increase in
$E_{\rm kin, av}$ that is responsible for the spectral evolution.
There is no significant indication of a supra-thermal tail in the electron
spectrum. 

\begin{figure}[htb]
\epsscale{1.178}
\plotone{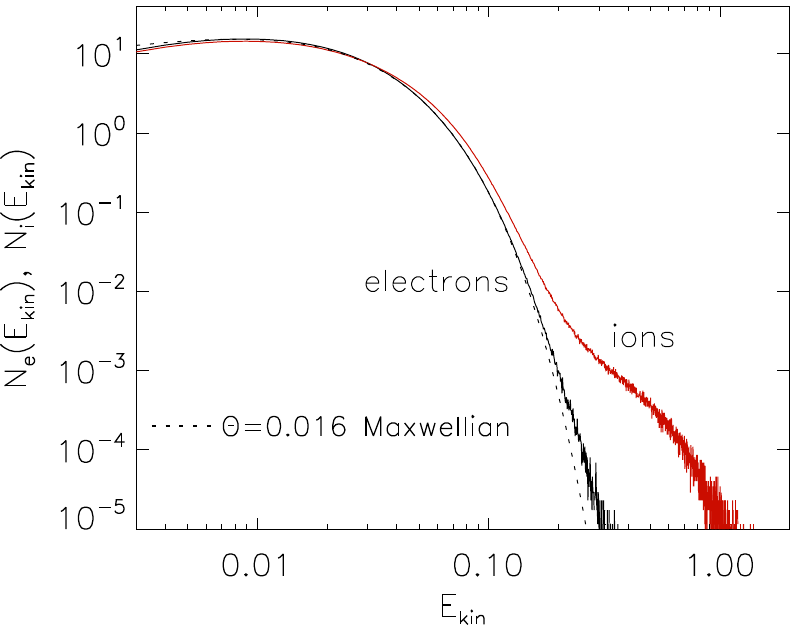}
\caption{Kinetic-energy spectra of dense electrons (black line) and dense ions
(red line) in the reverse-shock downstream region at $x/\lambda_{si}\approx
2850-2890$ and time 
$t=4111 \,\omega_{pi}^{-1}$ for the magnetized run M1, in the CD rest frame. The
spectra are normalized and expressed in simulation units, in which
$m_e\,c^2=0.25$. The dotted line shows a Maxwellian fit to the electron
spectrum.     
\label{mag9}}
\end{figure}

Dense-plasma particles downstream of the reverse shock are sampled in a slice of
width $\sim 40\, \lambda_{si}$, centered at $x/\lambda_{si}\approx 2870$ and
their
energy distributions in the CD rest frame at the end of run M1 are presented 
in Figure \ref{mag9}. The hydrodynamic plasma compression is fully established
in this narrow region at the very end of the simulation. The electrons and ions
carry to within 10\% the same kinetic energy. Only in the high-energy tail do we
observe a modest anisotropy in the electron distribution.
A Maxwellian fits the electron spectrum reasonably well; a slight discrepancy at
high energies coincides with anisotropy.  
The ions are anisotropic as well; a substantial population of returning
particles exists
that builds a superficial high-energy tail (see Fig. \ref{mag4}), which is
therefore not indicative of ion pre-acceleration. 

\section{SUMMARY AND DISCUSSION}\label{concl}

We have performed 2D3V PIC simulations of non-relativistic plasma collisions
with absent
{or} parallel large-scale magnetic field. Contrary to earlier expectations,
simulations
by \citet{kato08} suggested that Weibel-type instabilities can trigger shock
formation
in an {unmagnetized} environment even for collisions with non-relativistic velocity, but
fast enough to prohibit an electrostatic shock 
\citep{
1970PhRvL..25.1699F,1971PhRvL..27.1189F}. {A} pre-existing magnetic
field {may therefore be} not necessary for the formation of non-relativistic shocks. The
question {we addressed in this study is}
whether or not the shock structure and 
the particle spectra reach a steady state, and to what degree a parallel
magnetic field 
renders shock formation more efficient. A secondary goal was to investigate the
efficacy of
particle pre-acceleration, which appears needed for their injection into a
diffusive shock-acceleration process.

To address these questions we follow the evolution of
the system longer than was done by \citet{kato08}. Also, we are interested in
asymmetric flows, i.e.
the collision of plasma slabs of different density, leading to two different
shocks and a contact
discontinuity which is self-consistently modeled. 
Emphasis has been placed on setting up the simulations as clean as possible.

The {simulation} parameters are chosen to be close to those of young supernova remnants, except that
we use a reduced {ion-to-electron mass ratio}, $m_i/m_e=50$. About 50 years after the explosion of a type-Ia supernova, 
the density ratio between ejecta and ambient medium is about 10 as assumed here, 
{  if the density of the ambient gas is about 1 atom$/cm^3$.} 
The shock 
velocities are about $0.05\,c$, a factor of a few lower than in our simulations. In the magnetized
case, the Alfv\'{e}nic Mach number of the forward shock is $M_{\rm Af}=c\beta_{\rm
fR,M}/v_{\rm A,R}\simeq 21.1$ and that of the reverse shock $M_{\rm
Ar}=c\beta_{\rm rL,M}/v_{\rm A,L}\simeq 22.5$. Here, {  as forward shock we denote}
the shock 
propagating into the low-density {(ambient)} plasma; vice versa the reverse shock travels through
the dense plasma, {that represents the ejecta}.

Our results can be summarized as follows:
\begin{itemize}
\item
Both in the unmagnetized and the magnetized case, a {double}-shock structure builds on time scales of about a thousand ion plasma times, $\omega_{pi}^{-1}$. 
{The colliding plasmas are initially very cold, so that the sonic} Mach number in our simulations is high
enough to prevent an electrostatic shock. {We show that while non-relativistic shocks 
in both unmagnetized and magnetized plasmas can be mediated by Weibel-type
instabilities, the efficiency 
of shock-formation processes is higher {  when a large-scale magnetic field is present.} 
The amplitude of magnetic turbulence generated in the collision region is in the magnetized plasma an order of magnitude larger than in unmagnetized conditions. 
The typical magnetization of the {  ambient medium, into} which young SNR 
shocks propagate, given by $\sigma=M_{\rm A}^{-2}$, is in the range 
$10^{-6} < \sigma < 10^{-3}$. Our magnetized run with 
$\sigma\approx 2\times 10^{-3}$ thus probes an upper limit of the sigma 
parameter, and any realistic {  formation scenario of parallel shocks in SNRs should} 
show characteristics intermediate between the two cases studied here.}

\item
In the unmagnetized case, small-scale electrostatic 
and filamentation{-like} instabilities operate in parallel and heat {the} electrons. Eventually, 
strong fluctuations arise in the density of electrons and ions that lead to the formation 
of {the double-shock structure}. The electron dynamics play an important role in the development of the system.
Ion-ion or ion-electron streaming generally drives the turbulence, {which is mainly magnetic}.
The exact type of {an} instability, however,
depends on {the} location in the shock precursor and is generally different for forward and reverse shocks.
Nonlinearities and the change of conditions during {plasma} advection toward the shocks render an identification 
with linear modes difficult. In the magnetized case, filamentation {in the dense plasma region} is initially more pronounced, 
which we verify with linear analysis, but it quickly evolves into an oblique mode and nonlinear 
density fluctuations. Some {of} the streaming instabilities that lead to the formation of {the reverse shock} are
similar to non-resonant modes that have been discussed in the context of cosmic-ray induced 
magnetic-field amplification 
\citep[e.g.][]{Bell2004,niem08,2009ApJ...694..626R,2009ApJ...706...38S,2010ApJ...709.1148N,stroman}.
{In both the magnetized and unmagnetized systems, magnetic-field generation processes lead to stronger magnetic fields downstream of the forward shocks, rather than reverse shocks. We would thus expect synchrotron emission to originate from the magnetized structure of the forward shock transition, which is in agreement 
with SNR observations.}

\item The electron distributions {downstream of the forward and reverse shocks} 
are generally isotropic, whereas that
is not always the case for the ions. The observed spectra suggest that the electrons 
have reached a statistical equilibrium and generally carry nearly as much 
kinetic energy as do the ions, although the {latter} have not yet reached an
equilibrium.  

\item We do not see {any significant} evidence of
pre-acceleration, neither in the electron population nor in the ion
distribution. Note that the shock-surfing acceleration, that reportedly provides
fast electron acceleration \citep{2009ApJ...690..244A}, should not
operate in situations with absent or strictly parallel magnetic field, because 
${\bf v}\times {\bf B}=0$. 
In our magnetized simulation,
the Alfv\'enic Mach number is slightly below the threshold 
(at $M_{\rm A}\gtrsim 25$) {for} the electron-acceleration process described by
\citet{2010PhRvL.104r1102A}. \citet{2010A&A...509A..89D} and \citet{mur10} 
report efficient electron acceleration to
Lorentz factors of order 100 in PIC simulations of slightly faster plasma collisions {than} 
discussed here ($0.5 c$ and $0.9 c$ instead of $\sim 0.4 c$). In both cases the 
{strong ($\Omega_e/\omega_{pe}\sim 1$)} large-scale magnetic field is assumed to be obliquely oriented relative to the 
flow axis, which provides for more effective particle reflection in the perpendicular 
component of the magnetic field at the shock transitions. Being interested in
ion dynamics, \citet{2012ApJ...744...67G} present hybrid simulations and a PIC simulation 
of a parallel shock with $M_{\rm A}=4.7$, about 4.5 times the magnetization assumed
here. They find emergent ion acceleration very late in the PIC simulation
(at $t\simeq 87\,\Omega_i^{-1}$) and also during the subsequent evolution captured with their
hybrid simulations. They also find that ion acceleration proceeds to higher energies if
the Alfv\'enic Mach number is larger, {although the injection efficiency decreases}. In ion plasma times, our magnetized PIC
simulation has similar
duration as that of \citet{2012ApJ...744...67G}, but is nearly 4
times shorter in ion gyro-times on account of the lower {plasma} magnetization{ assumed in our simulations. However, the low Alfv\'enic Mach number used in the PIC simulation in \citet{2012ApJ...744...67G} {  suppresses
Weibel-type instabilities. They are also not modeled with their hybrid 
simulations for faster shocks (their $M_{\rm A}=31$ run should be comparable with ours).} Because we demonstate that electron dynamics are important for the evolution of the system, a direct comparison with these results is not possible. Nevertheless,}
all spectral
structures we find coincide with anisotropic distribution functions, suggesting {an} imperfect
{ion-}beam relaxation rather than acceleration per se. 

\item
{We have also studied the effect of the assumed ion-to-electron mass ratio on our results. 
We find that although the main characteristics of the long-time evolution of the systems
do not critically
depend on the mass ratio within the range studied, the detailed mechanisms of reaching a steady state might be modified
or even different if the assumed mass ratio is too low. The timescales for
particle-energy equilibration and the efficiency of particle pre-accelaration processes might be
overestimated in simulations that use a reduced ion-to-electron mass ratio.}

\end{itemize}

Computer resource limitations prohibit us to study the properties of the shock formation for close-to-realistic ion-to-electron mass ratios. 
Studies of the saturation of the linear phase of purely-transverse ion-driven filamentation instability indicate a very low efficiency of the conversion of the ion beam energy into 
magnetic-field energy for a realistic mass ratio \citep[e.g.,][]{ren07,wiersma04}, suggesting that other instabilities must mediate the shock formation in astrophysical settings. However, the applicability of these estimates to shocks might be limited, e.g., because they do not consider a nonlinear evolution of the filamentation instability and also not a steadily driven instability. In fact,
recent numerical simulations of relativistic collisions of the electron-ion plasmas demonstrate the formation of the filamentation-instability mediated shocks, whose properties do not significantly change with ion-to-electron mass ratio between 16 and 1000 
\citep{2008ApJ...673L..39S}. Since the characteristics of the magnetic-field generation in our simulations are analogous to these observed in relativistic shocks, we similarly expect the properties of Weibel-instability-mediated nonrelativistic shocks will not be substantially modified for a realistic mass ratio.

Concerning the energy equipartition between electron and ions, as well as the 
absence of significant non-thermal tails in the particle spectra,
the question arises to what degree 
residual 2-body collisions in the simulation provide thermalization that in the 
highly-collisionless space plasma would not be relevant. 
{In space environments} the mean free path for
collective plasma interactions is many orders of magnitude smaller than that
of 2-body collision. In PIC simulations one uses few computational particles to 
represent very many real electrons {or} ions and thus introduces artificial
collisionality. Would that impact, and possibly prevent,
particle pre-acceleration in our simulations?

Coulomb collisions occur with a minimum impact parameter of about 1 grid-cell 
length on account of the extended charge distribution used to represent the 
particle. There are always many ions and electrons within one grid cell, and 
therefore substantial shielding will occur. {Tests
indicate that 2-body Coulomb interactions are indeed negligible, but random
charge imbalances that arise from a low number of particle per cell are possibly not.
In fact, they can provide artificial scattering on time scales similar to or shorter
than the simulation time for electron velocities $\lesssim 0.03\,c$. The post-shock
electrons, in particular those in the spectral tail, are always considerably faster than
that, and therefore we conclude that in our simulations
artificial scattering can not suppress the formation 
of non-thermal tails in the electron spectrum.}

\acknowledgments
The authors thank Mark Dieckmann for comments and useful discussions.
The work of J.N. is supported
by Narodowe Centrum Nauki as research project DEC-2011/01/B/ST9/03183. 
M.P. and V.W. acknowledge support through grant PO 1508/1-1 of the Deutsche
Forschungsgemeinschaft.
The work of A.B. is supported by the project ENE2009-09276 of the Spanish
Ministerio de Educacion y Ciencia. Simulations have been performed at the
Pleiades facility at the
NASA Advanced Supercomputing (NAS).


\begin{appendix}

\section{SIMULATIONS WITH LOWER ION-ELECTRON MASS RATIO}
The present study is based on simulations that use a reduced ion-electron mass
ratio,
$m_i/m_e=50$. To ascertain the effect of this parameter choice on our results,
two additional 
large-scale numerical experiments with $m_i/m_e=20$ have been performed 
(run U2 and M2; see Table \ref{t1}).

We find that the main characteristics of the long-time evolution of the systems
do not critically
depend on the ion-electron mass ratio within the range studied. In both the
unmagnetized and
magnetized case, a system of forward and reverse shocks, separated by a CD, is
formed. 
The late-stage structure and nature of the electromagnetic turbulence is
analogous to that observed
for runs with $m_i/m_e=50$. Also, the processes of particle-energy dissipation
and 
re-distribution proceed qualitatively in the same fashion. 
However, the detailed mechanisms of reaching a steady state might be modified
or 
even different if the assumed mass ratio is too low. 
In the following, we only detail the most important modifications of the system
characteristics introduced by a lower mass ratio.  

\begin{figure*}[htb]
\begin{center}
\includegraphics[scale=1.25]{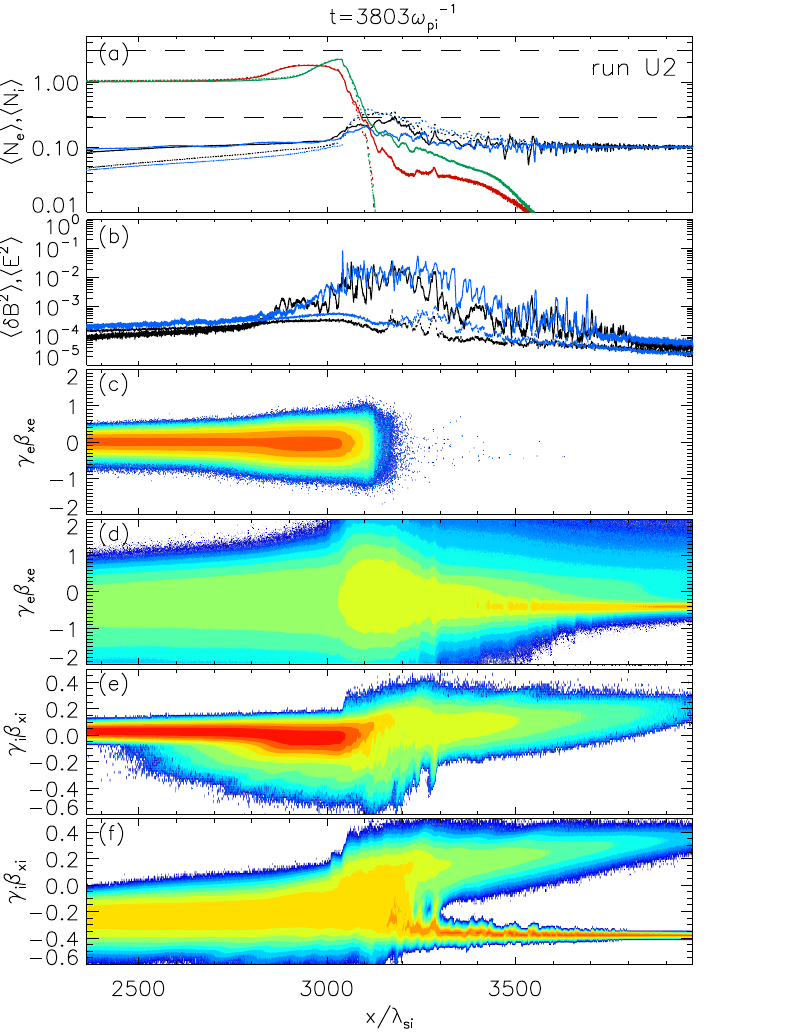}\hspace*{-2cm}
\includegraphics[scale=1.25]{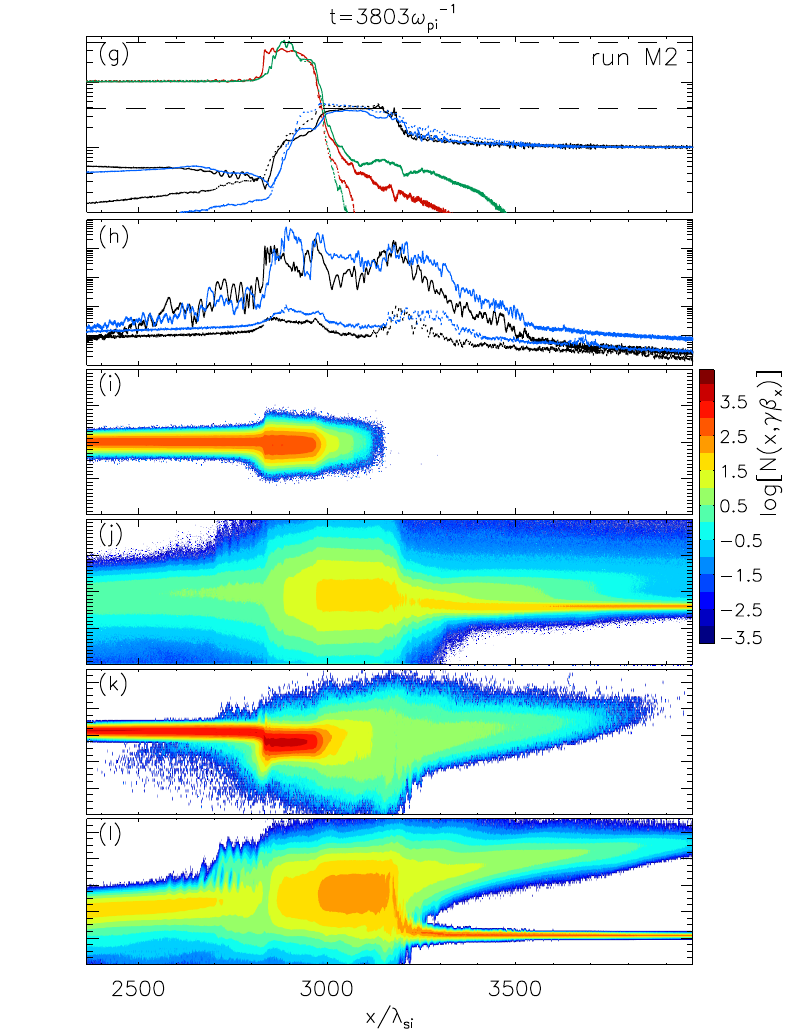}
\end{center}
\caption{Structure of the collision region at time $t=3803\, \omega_{pi}^{-1}$
for (left) unmagnetized run U2 and (right) magnetized run M2 {(see Fig.
\ref{unm3})}.
In Figs. \ref{app1}a and \ref{app1}g,
results for run U1/M1 are plotted for $t=3803\, \omega_{pi}^{-1}$ with green
lines (dense plasma) and blue lines (tenuous plasma) for comparison. 
{The horizontal dashed lines mark the hydrodynamic compression level for the
forward and reverse shocks (see Figs. \ref{unm3} and \ref{mag3}).}
Figs. \ref{app1}b and \ref{app1}h overplot the magnetic and electric
energy-density profiles for runs U1 and M1, respectively.
{The mean magnetic field with energy density of $7.9\times 10^{-3}$ is subtracted in the magnetic energy-density
profile in Fig. \ref{app1}h.}
\label{app1}}
\end{figure*}

\begin{figure}[htb]
\epsscale{0.59}
\plotone{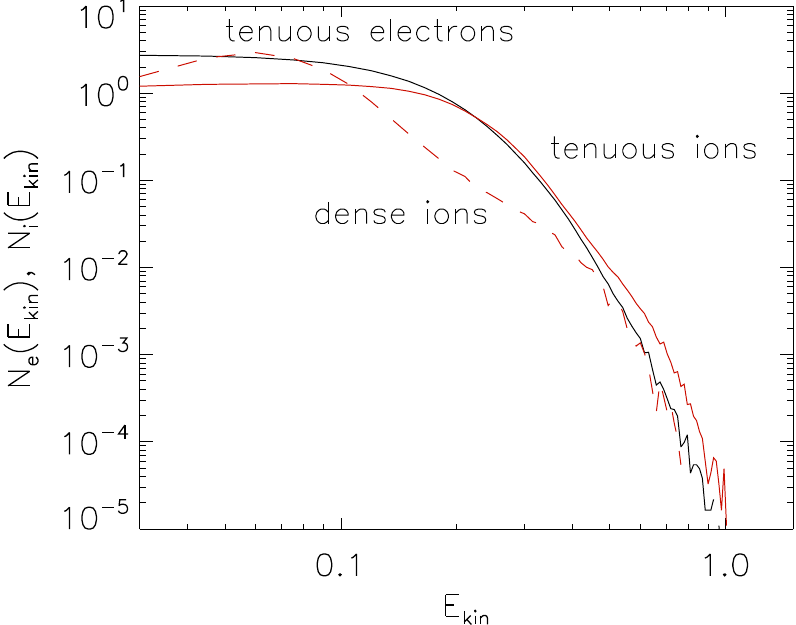}
\caption{{Normalized kinetic-energy spectra of tenuous electrons (black solid line),
tenuous ions (red solid line), and the dense ions (red dashed line) downstream
of the forward shock at
$x/\lambda_{si}\approx 3100$ and time $t=3803 \,\omega_{pi}^{-1}$ for the
unmagnetized run U2, in the CD rest frame. The ion-electron mass ratio is
$m_i/m_e=20$, and in simulation units $E_{\rm kin}=0.25\,(\gamma-1)$ for electrons
and $E_{\rm kin}=5\,(\gamma-1)$ for
ions.}
\label{app2}}
\end{figure}

\begin{figure}[htb]
\epsscale{1.17}
\plottwo{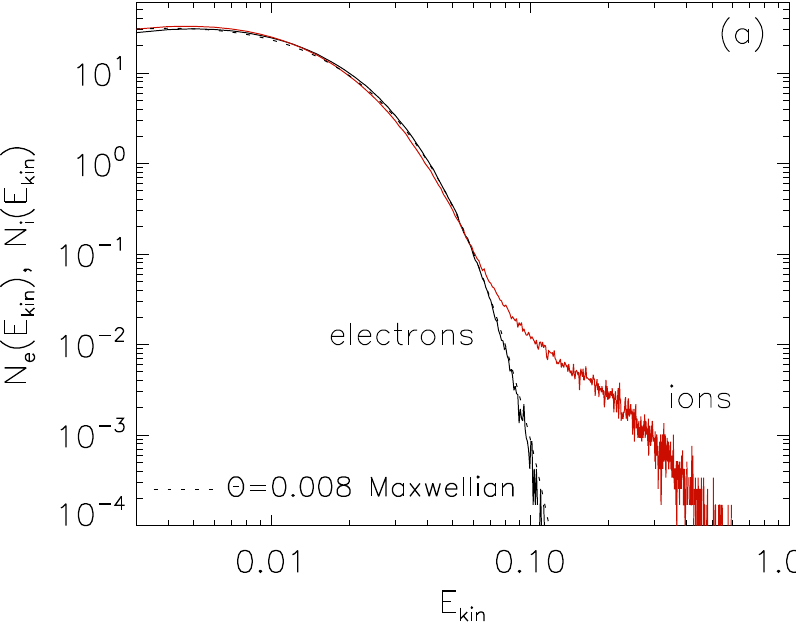}{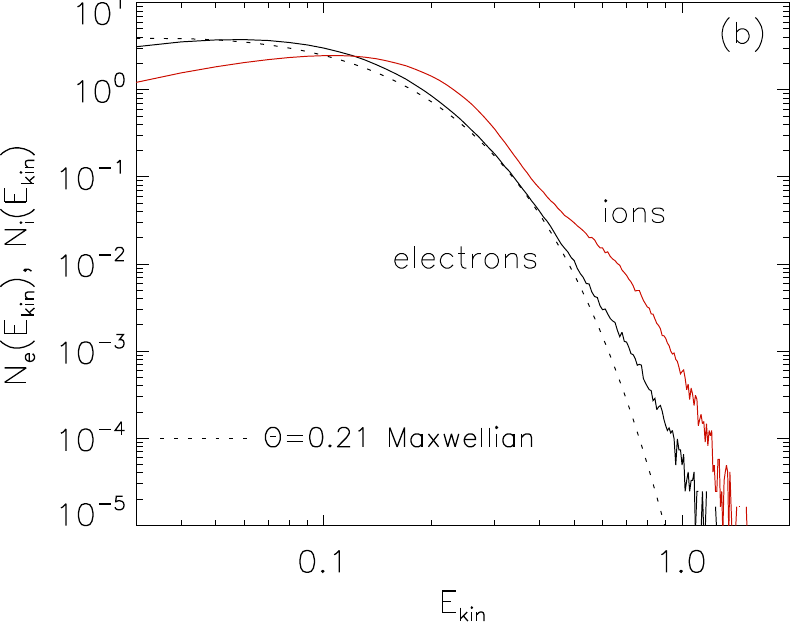}
\caption{(a) Kinetic-energy spectra of dense electrons (black line) and dense
ions
(red line) in the reverse-shock downstream region at $x/\lambda_{si}\approx
2850-2890$ and time $t=3803 \,\omega_{pi}^{-1}$ for the magnetized run M2. The
dotted line shows a Maxwellian fit to the electron
spectrum. (b) Kinetic-energy spectra of tenuous electrons (black line) and
tenuous
ions (red line) in the forward-shock downstream region at $x/\lambda_{si}\approx
3000$ and time $t=3803 \,\omega_{pi}^{-1}$ for the magnetized run M2. The dotted
line shows a relativistic Maxwellian for comparison. The spectra are shown in CD
rest frame and are normalized and expressed in simulation units, in which
$m_e\,c^2=0.25$.
\label{app3}}
\end{figure}

Figure \ref{app1} presents the structure of the collision region at the end of
run U2 (left) and 
M2 (right) at time  $t=3803 \,\omega_{pi}^{-1}$ (to be compared with Figs.
\ref{unm3} and \ref{mag4}). 
Overlaid on the average particle densities and electromagnetic-field profiles
in
Figures \ref{app1}(a-b) and \ref{app1}(g-h) are the profiles obtained for runs
U1 and M1. 
The horizontal axis is normalized to the skin depth of dense ions, which is 
$\lambda_{si}\approx 36.2$ for $m_i/m_e=20$. The particle spectra, measured in
the CD rest frame,
in the downstream region of the forward shock are shown in Figure \ref{app2} for
run U2.  
Figure \ref{app3} displays spectra in the downstream region of the reverse (a) 
the forward (b) shock for run M2.

\noindent
Unmagnetized simulation:
\begin{itemize}
\item The initial collision of the tenuous- and dense-plasma slabs leads to 
magnetic-field generation via Weibel-like instabilities, whose level is
similar in run U1 and U2. However, due to their lower mass, the ions are more 
easily deflected in the magnetic field. Consequently, the ion-isotropization
rate 
and the rate of kinetic-energy exchange with electrons are faster. Thus, in run
U2 the CD 
acquires its steady-state velocity earlier, which 
accounts for a shift of the average profiles for runs U1 and U2 in Figures
\ref{app1}a 
and \ref{app1}b. A substantial part of incoming beam particles is also reflected
towards 
their respective upstream. In result, the contribution of dense ions to the
forward-shock
compression is significantly smaller in run U2 (compare solid red and green
lines 
to solid black and blue lines in Fig. \ref{app1}a). However, also in this case
the 
tenuous-ion filaments do not fully disintegrate in the shock downstream. Note a
slight 
difference in the electron-compression profile right from the CD (dotted black
and blue 
line in Fig.~\ref{app1}a) for the two mass ratios. 

\item The forward-shock electromagnetic-turbulence structure in run U2 is
analogous to
that of run U1. The role of the dense-ion beam in magnetic-field generation via 
Weibel-like instabilities and in the amplification of the 
{parallel charge-density} 
perturbations is in run U2 shared more equally between the beam of dense
ions and the 
returning tenuous ions (compare Fig. \ref{app1}e-f and \ref{unm3}e-f). The
magnetic-turbulence 
level in the filamentation region of the shock precursor ($3200 \lesssim
x/\lambda_{si} \lesssim 3500$) 
is considerably lower for
$m_i/m_e=20$, which results from the higher beam temperatures that cause a 
serious {reduction} of the growth rate of filamentation-like modes. 

\item The tenuous-ion beam that permeates the dense-plasma region left from the
CD is much 
warmer in run U2 (compare Fig. \ref{app1}f and \ref{unm3}f). This causes a
decrease in the growth 
rate of the oblique filamentation modes and results in much weaker turbulent
precursor to the reverse shock 
in run U2 (Fig. \ref{app1}b). However, an increased efficiency of ion 
reflection in the CD region leads to an effective decoupling of the dense-ion
beam from the 
tenuous plasma; the numerous returning dense ions efficiently amplify the
magnetic turbulence 
in the reverse-shock transition. The detailed structure of the reverse shock in
run U2 is thus 
considerably modified with respect to that observed in run U1 (Fig.
\ref{app1}a-b for 
$x/\lambda_{si} \lesssim 3010$). However, also for mass ratio
$m_i/m_e=20$, the energy density of the magnetic field left of the CD
is more 
than an order of magnitude smaller than in the forward-shock region and the
reverse-shock 
compression does not reach the hydrodynamic limit.

\item Particle spectra downstream of the forward shock (calculated in a slice of
width
$\sim 20\, \lambda_{si}$, centered at $x/\lambda_{si}\approx 3100$, i.e.,
similar location 
as in run U1) are shown in Figure \ref{app2}. Whereas the electron and dense-ion
spectra
display characteristics analogous to those described for run U1, the tenuous-ion
distribution 
resembles that observed downstream of the forward shock in the {\it magnetized} 
simulation, M1. This includes a plateau section in the momentum distribution,
which 
only slightly deviates from isotropy in run U2. The contribution of tenuous ions
to the total 
kinetic energy is, at a level of 38\%, the same as in run U1. The dense ions
instead carry
17\% only; the difference of 8\% compared to run U1 is shifted to the electrons
which now 
account for 44\% of the total kinetic energy. These characteristics suggest that
the timescales
for particle-energy equilibration might be overestimated in lower-mass-ratio
simulations.
\end{itemize}

A slight dependence of the average density profiles on the ion-electron mass
ratio 
has been reported in \cite{kato08}. In our simulations, we see a similar effect
in the 
forward-shock profile. Even more significant modifications are observed in the 
region to the left from the CD. However, the reverse-shock formation has not
completed yet 
by the end of runs U1 and U2.

\noindent
Magnetized simulation:
\begin{itemize}
\item {A reduced mass ratio in run M2 results in larger Alfv\'{e}n
velocities, which for the tenuous and dense plasma now read, $v_{\rm
A,R}=2.9\cdot 10^{-2}c$ and $v_{\rm A,L}=8.8\cdot 10^{-3}c$, respectively. Thus
the simulation probes a slightly lower Alfv\'{e}nic-Mach-number limit
(but still $M_{\rm A}\gg 1$),
with Mach numbers $M_{\rm Af}\simeq 13.5$ and 
$M_{\rm Ar}\simeq 14.4$ {for the forward and reverse shock}, respectively.}  

\item The characteristic differences
introduced by a lower ion mass, {and hence larger rate of
ion isotropization,} 
that we described above for unmagnetized simulations, are also 
observed in the magnetized runs M1 and M2. 
{This suggests that it is the ion-beam temperature, and not the slight change
in the 
magnetic-field strength the ions experience through an increase of their
cyclotron frequency,
that is responsible for the observed differences in {the} development of Weibel-like
instabilities in both runs.}
The level of magnetic turbulence in the forward-shock 
precursor in run M1 considerably exceeds that in run M2 (Fig. \ref{app1}h). This
is because in
the magnetized case the 
turbulence is generated exclusively via filamentation-like instabilities, 
whose growth rates falls off with plasma temperature. Note, that the magnetic
field decays faster 
downstream of the forward shock in run M2.

\item The effects of the warm beams on the system evolution are even more
pronounced
in the region left from the CD. Instead of the purely-transverse
filamentation-like mode excited 
in the dense-plasma region by an energetic tenuous-ion beam in run M1 (see Sec.
\ref{emag}), a 
much weaker oblique mode is generated in simulation M2. Consequently, the
current filaments 
are weak and their merging does not lead to the strongly-nonlinear "magnetic
bubble" effects that 
in run M1 were largely responsible for the efficient decoupling of the colliding
ion beams. 
Nevertheless, also in this case the system can form the reverse-shock transition
at an
early stage. The shock formation is now mediated by a strong filamentation-like
instability 
due to the large number of dense ions reflected in the collision region.  

\item As in run M1, the formation of the reverse-shock transition at late stage
in run M2 involves 
the amplification and compression of {a} parallel magnetic Bell-like mode. These
processes are 
responsible for the turbulent-field structure at $x/\lambda_{si} \lesssim 2900$
in
Figure \ref{app1}h. The parallel left-hand circularly polarized mode has
properties similar to 
these observed in run M1: the wavelength 
$\lambda\sim 20\,\lambda_{si}$, frequency $\omega\simeq 0.43\,\Omega_i$, and the
phase velocity
$v_{ph}\simeq 1.4\, v_{\rm A,L}$. Note, however, that the reverse-shock
compression in run M2 does not 
reach the hydrodynamic limit by the end of the simulation. 

\item Particle spectra shown in Figure \ref{app3} display characteristics
similar to the 
distributions derived for run M1. 
{However, as is the case of run U2, the energy-equilibration
processes proceed with a slightly faster rate for a lower mass ratio. The
kinetic-energy content of the dense electrons
downstream of the reverse shock is in run M2 {the same} within 0.3\% as that of
the dense ions,
and the electrons are isotropic and thermal (compare a Maxwellian fit to the
electron spectrum 
in Fig. \ref{app3}a). Downstream of the forward shock, the tenuous electrons
carry 45\% of the total kinetic energy. An analysis of the time evolution of the
electron spectra shows an
indication of supra-thermal tails, that were not observed in run M1.
Therefore, although the electron spectra in run M2 may be influenced by low
statistics at high energies, this result demonstrates that
the efficiency of particle injection/pre-accelaration processes might be
overestimated in simulations that use a reduced ion-to-electron mass ratio.}
\end{itemize}


\section{DIELECTRIC TENSOR ELEMENTS}\label{ap:tensor}
The elements of the dielectric tensor (\ref{tensor}) read,
\begin{eqnarray}
a&=&1-\frac{1+R}{(x-Z_x)^2}-\frac{(a_4+a_5+a_3 R) \beta_1^2}{(x \beta_1+Z_x
\beta_2)^2},\\
b&=&1-\frac{Z_x^2}{x^2 \beta_1^2}-\frac{R}{x^2}
\left[\frac{(x-Z_x)^2/R}{(x-Z_x)^2-\Omega_B^2}+\frac{(x-Z_x)^2}{(x-Z_x)^2-R^2
\Omega_B^2}-B\right],\\
c&=&\frac{i R^2 \Omega_B }{ x^2}\left[\frac{(x-Z_x)/R^2}{
(x-Z_x)^2-\Omega_B^2}-\frac{x-Z_x}{(x-Z_x)^2-R^2 \Omega_B^2}+C\right],
\end{eqnarray}
with
\begin{eqnarray}
B&=&\frac{(a_4+a_5+a_3 R) (x \beta_1+Z_x \beta_2)^4-R (a_3+(a_4+a_5) R)
\beta_1^2 (x \beta_1+Z_x \beta_2)^2 \Omega_B^2}{R \left(1+R^2\right) \beta_1^2
(x \beta_1+Z_x \beta_2)^2 \Omega_B^2-R^3 \beta_1^4 \Omega_B^4-R (x \beta_1+Z_x
\beta_2)^4},\\
C&=&=\frac{(a_4+a_5)\beta_1 (x \beta_1+Z_x \beta_2)/R^2 }{ (x \beta_1+Z_x
\beta_2)^2-\beta_1^2 \Omega_B^2}-\frac{a_3 \beta_1 (x \beta_1+Z_x \beta_2) }{(x
\beta_1+Z_x \beta_2)^2-R^2 \beta_1^2 \Omega_B^2}.
\end{eqnarray}
In these expressions, $Z_x=k_x v_L/\omega_{pe,L}$ is the reduced flow-aligned
wave vector, $R$ is 
the electron-to-ion mass ratio, and $\Omega_B=\Omega_e/\omega_{pe,L}$ the
reduced electron cyclotron 
frequency.
The parallel magnetic mode appears in the reverse-shock precursor in the
late stage of our magnetized simulations M1 and M2 (Fig. \ref{mag6a}). However,
the mode can also be observed in the linear spectrum of the early-stage
turbulence (Fig. \ref{mag1a}). {We provide parameters
$a_n$ and $\beta_n$ for both setups in Table~\ref{t2}.}

\begin{deluxetable*}{lccccc}
\tablecaption{Parameters $a_n$ and $\beta_n$ of the dielectric tensor for the setups of Figures \ref{mag1a} and \ref{mag6a}; run M1. \label{t2}}
\tablewidth{0pt}
\tablehead{
Figure        &   $a_3$  &   $a_4$  &  $a_5$  &  $\beta_1$  &  $\beta_2$}
\startdata
\ref{mag1a}  &   0.1    &  0.01    &  0.09   &  0.0354     &   0.354     \\
\ref{mag6a}  &   0.05   &  0.005   &  0.045  &  0.0354     &   0.3    \enddata
\end{deluxetable*}

\end{appendix}


\end{document}